\documentclass[fleqn,usenatbib,usedcolumn]{mnras}
\usepackage[british]{babel}             
\usepackage{newtxtext}                  
\usepackage[slantedGreek]{newtxmath}    
\usepackage[T1]{fontenc}                
\usepackage{graphicx}                   
\usepackage{rotating}
\usepackage{dcolumn}
\newcolumntype{d}{D{.}{.}{-1}}
\def\dhead#1{\multicolumn{1}{c}{#1}}
\def\twolines#1#2{$\kern-6pt\Big\{ {\textrm{#1\hfill}\atop\textrm{#2\hfill}}$}
\def\vpad{{\Large$\mathstrut$}}
\usepackage{multirow}
\usepackage{hyperref}
\graphicspath{{.}{./FIGS/}{..}}

\defcitealias{2013MNRAS.429.2080W}{Paper I}
\defcitealias{2015MNRAS.453.4244W}{Paper II}
\defcitealias{2016MNRAS.462.2122W}{Paper III}

\hypersetup{pdfauthor={I. H. Whittam, D. A. Green, M. J. Jarvis and J. M. Riley},
               pdftitle={The faint radio source population at 15.7 GHz -- IV. The dominance of core emission in faint radio galaxies},
               pdfkeywords={galaxies: active, radio continuum: galaxies, catalogues, surveys},
               bookmarksnumbered=true}
\hypersetup{colorlinks=true,
            linkcolor=blue,
            citecolor=blue,
            filecolor=blue,
            urlcolor=blue}
\volume{{\rm in prep.}}
\title[Core emission in faint 15-GHz radio galaxies]{The faint radio source population at 15.7 GHz -- IV. The dominance of core emission in faint radio galaxies}

\author[I.~H.~Whittam et al.]{I.~H.~Whittam$^{1,2}$\thanks{imogen.whittam@physics.ox.ac.uk}, D.~A.~Green$^3$, M. J. Jarvis$^{1,2}$ and J.~M.~Riley$^3$\\
   $^{1}$Department of Physics and Astronomy, University of the Western Cape, Robert Sobukwe Road, Bellville 7535, South Africa\\
   $^{2}$Astrophysics, University of Oxford, Denys Wilkinson Building, Keble Road, Oxford, OX1 3RH\\
   $^{3}$Astrophysics Group, Cavendish Laboratory, 19 J.~J.~Thomson Avenue, Cambridge CB3 0HE}

\date{Accepted ---; received ---; in original form ---}

\pagerange{\pageref{firstpage}--\pageref{lastpage}}

\pubyear{2020}
\begin{document}

\label{firstpage}

\maketitle

\begin{abstract}
We present 15-GHz Karl G. Jansky Very Large Array observations of a complete sample of radio galaxies selected at 15.7 GHz from the Tenth Cambridge (10C) survey. 67 out of the 95 sources (71 per cent) are unresolved in the new observations and lower-frequency radio observations, placing an upper limit on their angular size of $\sim2$~arcsec. Thus compact radio galaxies, or radio galaxies with very faint jets, are the dominant population in the 10C survey. This provides support for the suggestion in our previous work that low-luminosity ($L<10^{25} \, \textrm{W Hz}^{-1}$) radio galaxies are core-dominated, although higher-resolution observations are required to confirm this directly. The 10C sample of compact, high-frequency selected radio galaxies is a mixture of high-excitation and low-excitation radio galaxies and displays a range of radio spectral shapes, demonstrating that they are a mixed population of objects.
\end{abstract}

\begin{keywords}
galaxies: active -- radio continuum: galaxies -- catalogues -- surveys
\end{keywords}

\section{Introduction}\label{section:intro}

Most deep studies of the extragalactic radio sky have focussed on frequencies around 1~GHz, due to the increased telescope time required to conduct a survey at higher frequencies, leaving the faint, high-frequency sky as a relatively unexplored parameter space. One major exception to this is the Tenth Cambridge survey (10C; \citealt{2011MNRAS.415.2708D,2011MNRAS.415.2699F,2016MNRAS.457.1496W}) at 15.7~GHz which is complete to 0.5~mJy in ten different fields, and extended to 0.1~mJy in two of these fields, making it the deepest high-frequency radio survey to date. 

In the first three papers in this series (\citealt{2013MNRAS.429.2080W,2015MNRAS.453.4244W,2016MNRAS.462.2122W}, hereafter \citetalias{2013MNRAS.429.2080W}, \textcolor{blue}{II} and \textcolor{blue}{III} respectively) we studied the properties of a complete sample of faint ($\sim 1$~mJy) sources selected at 15.7 GHz from the 10C survey in the Lockman Hole. We have used a combination of observations at other radio frequencies and data from across the electromagnetic spectrum to extensively study the properties of these sources and found that they are essentially all ($>94$ per cent) radio-loud AGN with a median redshift $z \sim 0.9$ \citepalias{2015MNRAS.453.4244W}. While the brighter sources in this sample have steep radio spectra, as expected, the fainter radio galaxies have flatter spectra than previously assumed; the median spectral index\footnote{The convention $S \propto \nu^{-\alpha}$, for flux density $S$, frequency $\nu$ and spectral index $\alpha$, is used throughout this work.} between 15.7 GHz and 610 MHz changes from 0.75 for sources with $S_{15~\rm GHz} > 1.5$~mJy to 0.08 for $S_{15~\rm GHz} < 0.8$~mJy, \citepalias{2013MNRAS.429.2080W}. This suggests that these fainter radio galaxies are dominated by emission from the core, which generally has a flatter radio spectrum (due to the super-position of several synchrotron self-absorbed components from the base of the jet) than the optically thin synchrotron emission from any extended lobes. Some of the sources display extended morphologies typical of Fanaroff and Riley type I or II sources (FRI or FRII; \citealt{1974MNRAS.167P..31F}), based on 610-MHz GMRT observations with a resolution of 5~arcsec \citepalias{2016MNRAS.462.2122W}. The majority (68 out of 96, 71 per cent), however, appear to be compact radio galaxies \citepalias{2016MNRAS.462.2122W}, which are referred to as `FR0' sources (the FR0 classification was introduced by \citealt{2011AIPC.1381..180G} to describe the class of weak, compact radio sources described by \citealt{2009A&A...508..603B}, but also studied by \citealt{1994MNRAS.269..928S} and possibly even earlier). The sources are a mixture of high-excitation and low-excitation radio galaxies (HERGs and LERGs, see e.g.\ \citealt{2012MNRAS.421.1569B}), and the HERGs tend to have flatter spectra and be more core-dominated than the LERGs \citepalias{2016MNRAS.462.2122W}.

Most models of the high-frequency extragalactic radio sky (e.g.\ \citealt{2005A&A...431..893D,2008MNRAS.388.1335W,2011A&A...533A..57T}) are based on extrapolations from lower frequencies, and are a poor fit to the observed 15-GHz source counts below $\lesssim$10~mJy. One such model is the SKA Simulated Skies (\citealt{2008MNRAS.388.1335W,2010MNRAS.405..447W}, S$^3$), which is a set of simulations of the radio and sub-mm universe. One component is a semi-analytic simulation of a $20 \times 20$~deg$^2$ patch of the extragalactic sky out to a redshift of 20, containing 320 million radio components.
While this simulation reproduces the observed source distributions at lower frequencies, it under-predicts the number of sources observed in the Ninth Cambridge (9C; \citealt{2003MNRAS.342..915W,2010MNRAS.404.1005W}) and 10C surveys with $S_{18~\rm GHz} < 10$~mJy by a factor of two, and fails to reproduce the observed spectral index distribution. In \citet{2017MNRAS.471..908W} we showed that by applying a simple modification to the simulation and assuming that the cores of the fainter FRI sources in the simulation are more dominant than previously thought, the observed high-frequency source counts can be reproduced. The observations are best matched by assuming that the fraction of the total 1.4-GHz flux density that originates from the core varies with 1.4-GHz luminosity; sources with 1.4-GHz luminosities $<10^{25}~\rm W \, Hz^{-1}$ require a core fraction $S_\textrm{core}/S_\textrm{tot} \sim 0.3$, while the more luminous sources require a much smaller core fraction of $5 \times 10^{-4}$. The more recent Tiered Radio Extragalactic Continuum Simulation (T-RECS; \citealt{2019MNRAS.482....2B}) models steep spectrum sources, flat-spectrum radio quasars and BL Lacs separately and is a significantly better fit to the observed high-frequency source counts. The primary reason for the better fit to the data is the inclusion of a significant population of low-powered radio-loud AGN with flat spectra, which are present in the observed 10C sample but not included in S$^3$. This is discussed further in Section~\ref{section:S3}.

In this paper we describe new, higher resolution, observations of a sub-sample of 10C sources made with the Karl G. Jansky Very Large Array (VLA). The observations were made at 15 GHz in C-configuration, so are matched in frequency to the original 10C observations but have a significantly higher resolution of $\sim$2~arcsec compared to 30 arcsec for 10C. They allow us to constrain the fraction of the total flux density originating from the cores of the radio galaxies, and therefore test our proposed modifications to S$^3$.

These new observations also allow us to further investigate the properties of this unique sample of high-frequency selected faint, compact radio galaxies. Despite the fact that they dominate lower-powered radio AGN samples (e.g.\ \citealt{2014MNRAS.438..796S,2018A&A...609A...1B}), the nature of compact radio galaxies (FR0 sources) is not well understood. While some compact radio galaxies are young sources that may go on to become more extended FRI or FRII galaxies, the number density of compact sources compared to that of extended sources implies that most must be older sources that have failed to produce large extended structures \citep{2014MNRAS.438..796S,2018A&A...609A...1B}. \citet{2015A&A...576A..38B} found no difference in the properties of the host galaxies (e.g.\ optical morphology, colour, stellar mass) of FRI sources compared to the more compact FR0 sources, although there is more recent evidence that FR0 sources are hosted by less massive galaxies than FRIs \citep{2017MNRAS.466.4346M,2018A&A...609A...1B}. The reason for the lack of extended emission in some radio galaxies is therefore not clear; possibilities include intermittent activity of the central engine \citep{1997ApJ...487L.135R,2002NewAR..46..263C,2009ApJ...698..840C}, jets which are easily interrupted due to a lower bulk speed or interactions with surrounding gas (e.g.\ \citealt{2015A&A...576A..38B,2016AN....337..105S,2018MNRAS.476.5535T,2019MNRAS.482.2294B}).

In Section~\ref{section:sample} we outline the sample selection and the radio and multi-wavelength data that are already available. Section~\ref{section:obs_data} describes the new VLA observations and data reduction and in Section~\ref{section:opt-matching} we update the multi-wavelength counterparts associated with the radio sources based on these observations. The new observations are compared to the 10C observations in Section~\ref{section:10C}. In Section~\ref{section:S3} we examine our measurements of radio core dominance in relation to the modification to S$^3$ we suggested in \citet{2017MNRAS.471..908W} and the T-RECS simulation, and discuss what these observations tell us about the properties of faint, compact radio galaxies. We then present our conclusions in Section~\ref{section:conclusions}.

Throughout this paper, we assume  $H_0 = 70~\textrm{km s}^{-1} \textrm{Mpc}^{-1}$, $\Omegaup_\textrm{M} = 0.3$, and $\Omegaup_{\lambda} = 0.7$ and use J2000 coordinates.

\begin{figure}
\centerline{\includegraphics[width=\columnwidth]{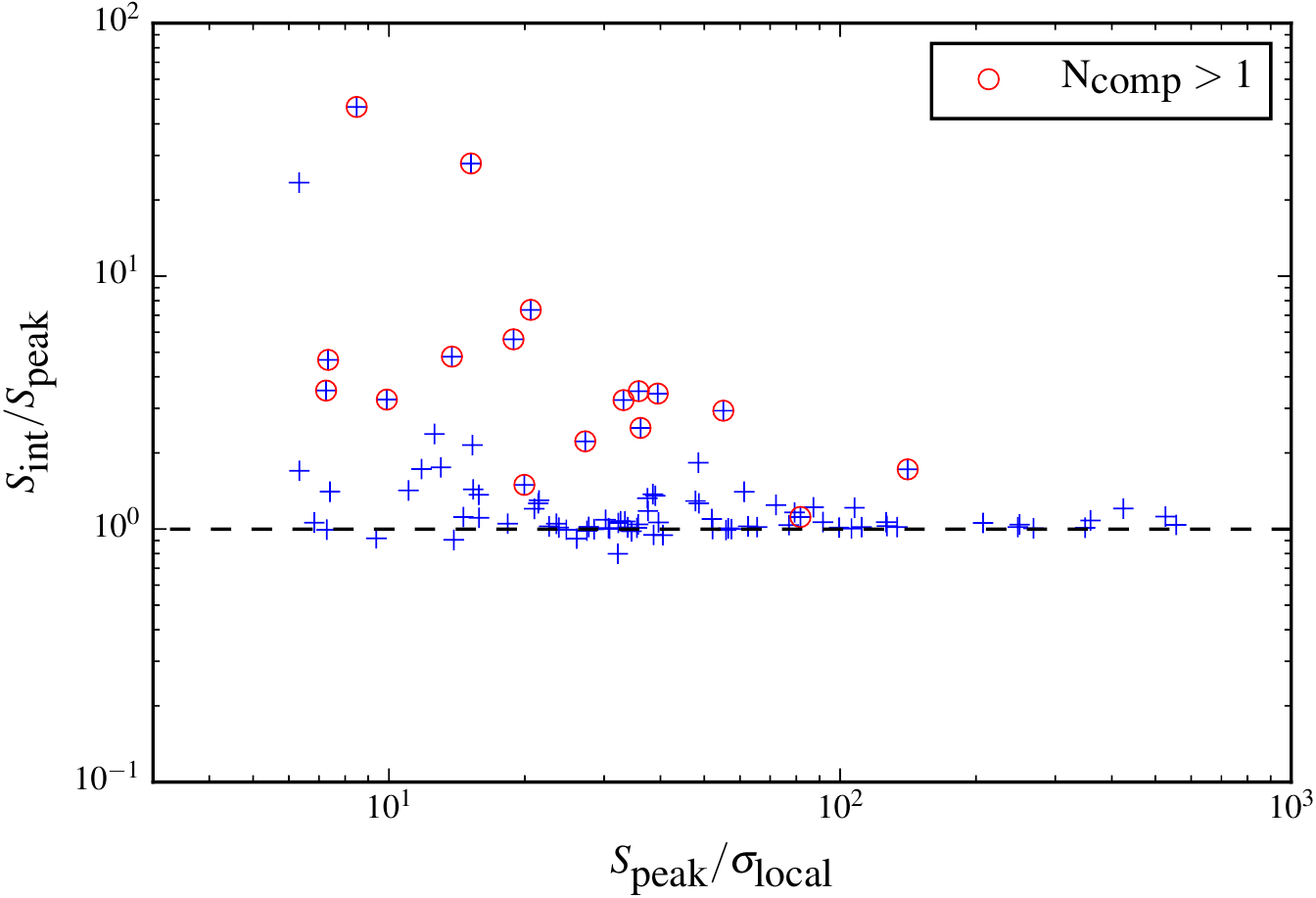}}
\caption{Ratio of the integrated to peak flux densities as a function of signal to noise ($\sigma_\textrm{local} = $ local rms noise). The red circles are sources with more than one component.}\label{fig:SpeakSint}
\end{figure}

\section{Sample used}\label{section:sample}

This work is based on a sample selected from the 10C survey at 15.7~GHz. The 10C survey was observed with the Arcminute Microkelvin Imager (AMI; \citealt{2008MNRAS.391.1545Z}) Large Array and covers 27 deg$^2$ complete to 1~mJy in ten different fields, and a further 12 deg$^2$ complete to 0.5~mJy contained within these fields. Here we use a complete sample of 96 sources selected from a region of the 10C survey in the Lockman Hole with particularly good low frequency data available. These 96 sources form a subsample of the full 10C Lockman Hole sample discussed in \citetalias{2013MNRAS.429.2080W}; full details of the sample selection are given in \citetalias{2015MNRAS.453.4244W}.

In \citetalias{2013MNRAS.429.2080W} the sample was matched to several lower-frequency (and higher-resolution) radio catalogues available in the field. These are: a deep 610~MHz Giant Meterwave Radio Telescope (GMRT) image \citep{2008MNRAS.387.1037G,2010BASI...38..103G}, a 1.4~GHz Westerbork Synthesis Radio Telescope (WSRT) image \citep{2012rsri.confE..22G,2018MNRAS.481.4548P}, two deep Very Large Array (VLA) images at 1.4~GHz which only cover part of the field \citep{2006MNRAS.371..963B,2008AJ....136.1889O}, the National Radio Astronomy Observatory (NRAO) VLA Sky Survey (NVSS; \citealt{1998AJ....115.1693C}) and the Faint Images of the Radio Sky at Twenty centimetres (FIRST; \citealt{1997ApJ...475..479W}). All but one of the 96 sources in this sample have a match in at least one of the lower-frequency catalogues, allowing radio spectral indices and sizes to be found for these sources. 

There is a wealth of multi-wavelength data available in the Lockman Hole, most of which is contained in the `SERVS Data Fusion' catalogue \citep{2015fers.confE..27V}, a multi-wavelength catalogue selected in the mid-infrared. In the Lockman Hole this catalogue contains data from the \emph{Spitzer} Wide-Area Infrared Extragalactic survey (SWIRE; see \citealt{2003PASP..115..897L}), the \emph{Spitzer} Extragalactic Representative Volume Survey (SERVS; see \citealt{2012PASP..124..714M}) and the United Kingdom Infrared Telescope (UKIRT) Infrared Deep Sky Survey (UKIDSS; see \citealt{2007MNRAS.379.1599L}) with deep optical photometric data taken by \citet{2011MNRAS.416..927G} at the Isaac Newton Telescope (INT) and the Kitt Peak National Observatory (KPNO). In order to be included in the catalogue, a source must be detected in either the \emph{Spitzer} IRAC1 or IRAC2 bands. In \citetalias{2015MNRAS.453.4244W} the radio sources in the sample were matched to the SERVS Data Fusion catalogue; these matches are reviewed in Section~\ref{section:opt-matching} with the addition of the new, higher-resolution, radio data presented here.

There are also two X-ray surveys which each cover part of this field; one with \emph{Chandra} \citep{2009ApJS..185..433W} which covers $0.7~\rm{deg}^2$ and one with \emph{XMM-Newton} \citep{2008A&A...479..283B} which covers $\sim0.2~\rm{deg}^2$. The 10C sources are matched to these two catalogues in \citetalias{2016MNRAS.462.2122W}.

The 96 sources in this sample have 15.7-GHz flux densities in the range $0.5 < S_\textrm{10C} / \rm mJy < 45$ and a redshift distribution peaking at $z \sim 1$ \citepalias{2015MNRAS.453.4244W}. This sample probes a wide luminosity range $10^{21} < L_{1.4~\rm GHz} / \rm W \, Hz^{-1} < 10^{28}$.

\section{Observations and data processing}\label{section:obs_data}

\subsection{The VLA observations}

The new data\footnote{Project code: 17A-183} were taken with the Karl G. Jansky Very Large Array (VLA) in C configuration in the Ku band. Pointed observations were made of the 96 sources in the sample described in Section~\ref{section:sample}. The 75 sources with $S_\textrm{10C} > 0.75$~mJy ($S_\textrm{10C}$ is the 15~GHz peak flux density from the 10C catalogue) were observed for 50 seconds on source. The fainter sources with $S_\textrm{10C} < 0.75$~mJy (21 sources) were observed for a time scaled according to the 10C peak flux density of each source, to allow the fainter sources to be observed for longer. The time for each source was calculated to achieve an rms noise level of $S_\textrm{10C} / 25$. For example the faintest source in the sample has $S_\textrm{10C} = 0.35$~mJy and in order to achieve the required rms noise level of 14$~\muup$Jy/beam it was observed for 187 seconds. 

The observations were split into two scheduling blocks (SBs) containing 42 and 54 sources, and these were observed on 31 May 2017 and 11 June 2017 respectively. 3C286 was observed once at the start of each SB for flux density and bandpass calibration. J1035+5628 was observed approximately every 15 minutes for phase calibration. The Ku band receiver was used with a central frequency of 15 GHz and 48 $\times$ 128 MHz spectral windows in full polarization. 

The data were initially processed using the NRAO VLA pipeline\footnote{\url{https://science.nrao.edu/facilities/vla/data-processing/pipeline}}, which consists of a series of \textsc{casa}\footnote{\url{http://casa.nrao.edu/}} commands. The pipeline performs some initial flagging; this includes the removal of the initial few integration points of a scan where the antennas may not all be on source, the removal of data from shadowed antennas and automatic radio frequency interference (RFI) flagging. The pipeline then performs delay and bandpass corrections.

After the execution of the NRAO pipeline the observations were averaged using 8 second integration times. The sources were imaged using the \textsc{casa} task \textsc{clean}. For the five sources with significant extended emission clean boxes were placed by hand; the remaining sources were imaged automatically using 500 iterations, with cyclefactor = 5. Sources with a signal-to-noise ratio $ > 20$ were self-calibrated. Two rounds of self-calibration were performed, both phase-only, with a first solution interval of 60 seconds and a second of 30 seconds. In some cases errors were reported by the \textsc{casa} task \textsc{gaincal} due to insufficient signal to noise; for these sources the non-self-calibrated images were used. The signal-to-noise ratio in each image was compared after each stage of self calibration; if it went down at any point then the non-self-calibrated images were used. The final resolution is $\approx 1.6 \times 2.4$~arcsec$^2$ for sources observed in the first SB and $\approx 1.6 \times 4.0$~arcsec$^2$ for those observed in the second SB. The resolutions were different for the two SBs as they were observed at different elevations. The nominal largest scale structure for the VLA in C array at 15 GHz is 97~arcsec, but as these are snapshot observations the largest angular scale they are sensitive to is around 50~arcsec.

\subsection{Source properties}\label{section:properties}

The flux density of each source was estimated by fitting one or more Gaussians to each source using the \textsc{casa} task \textsc{imfit}. 22 sources were resolved by these observations. 17 of these are significantly extended so were poorly fit by a Gaussian; the integrated flux densities of these sources were found interactively using the \textsc{casa} task \textsc{imview}. In cases where the core and extended emission could be distinguished from each other, the flux density of the core was also found separately. The ratio of integrated to peak flux densities is shown in Fig.~\ref{fig:SpeakSint} as a function of signal to noise; it is clear that many of the sources in this sample are unresolved on arcsec scales. The 17 sources resolved into more than one component are indicated by red circles in Fig.~\ref{fig:SpeakSint}.

\begin{figure}
\centerline{\includegraphics[width=\columnwidth]{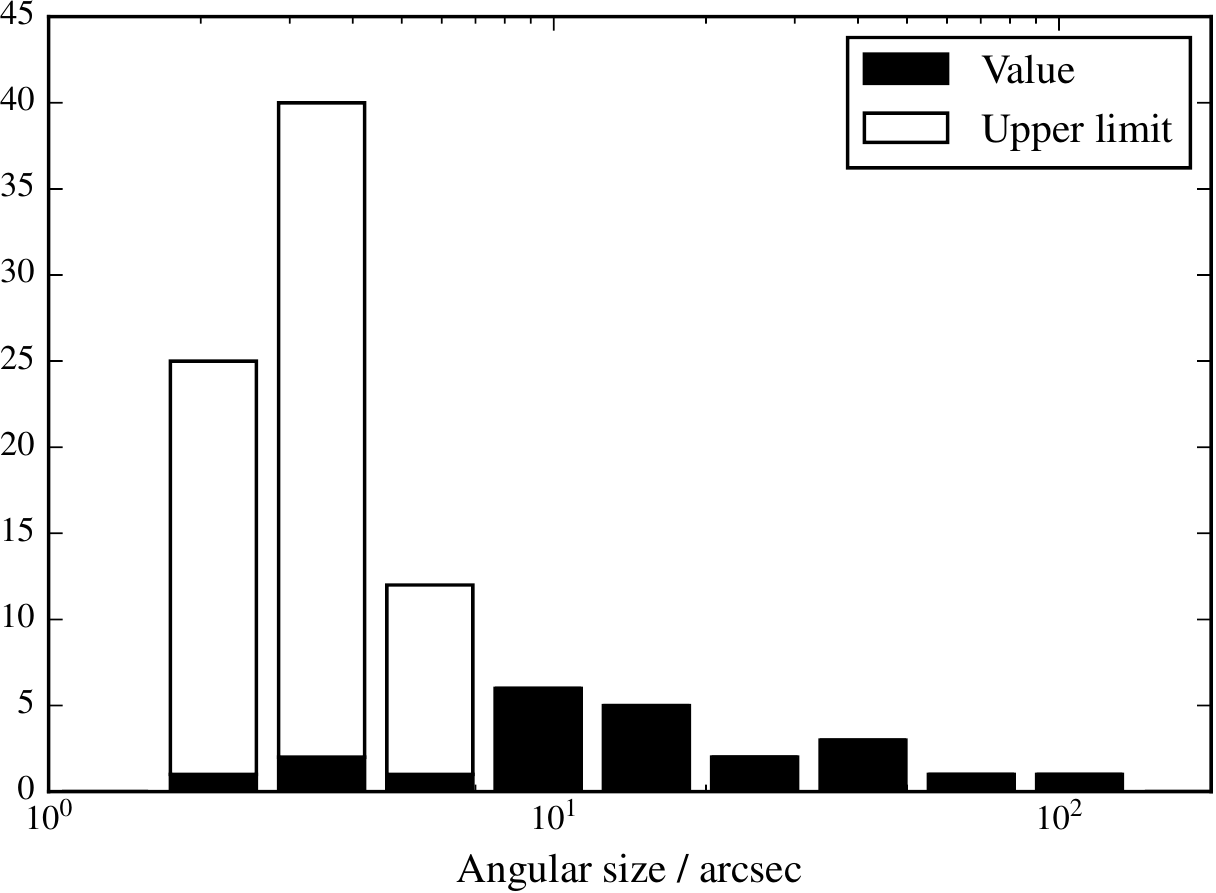}}
\caption{Distribution of angular sizes measured from the 15 GHz VLA images for all the sources in the sample. Upper limits are shown in white. Note that six sources which are unresolved in these observations have extended emission visible in 1.4~GHz and/or 610~MHz observations and are therefore not genuinely compact.}\label{fig:ang_size}
\end{figure}

\begin{figure*}
\centerline{\includegraphics[width=6cm]{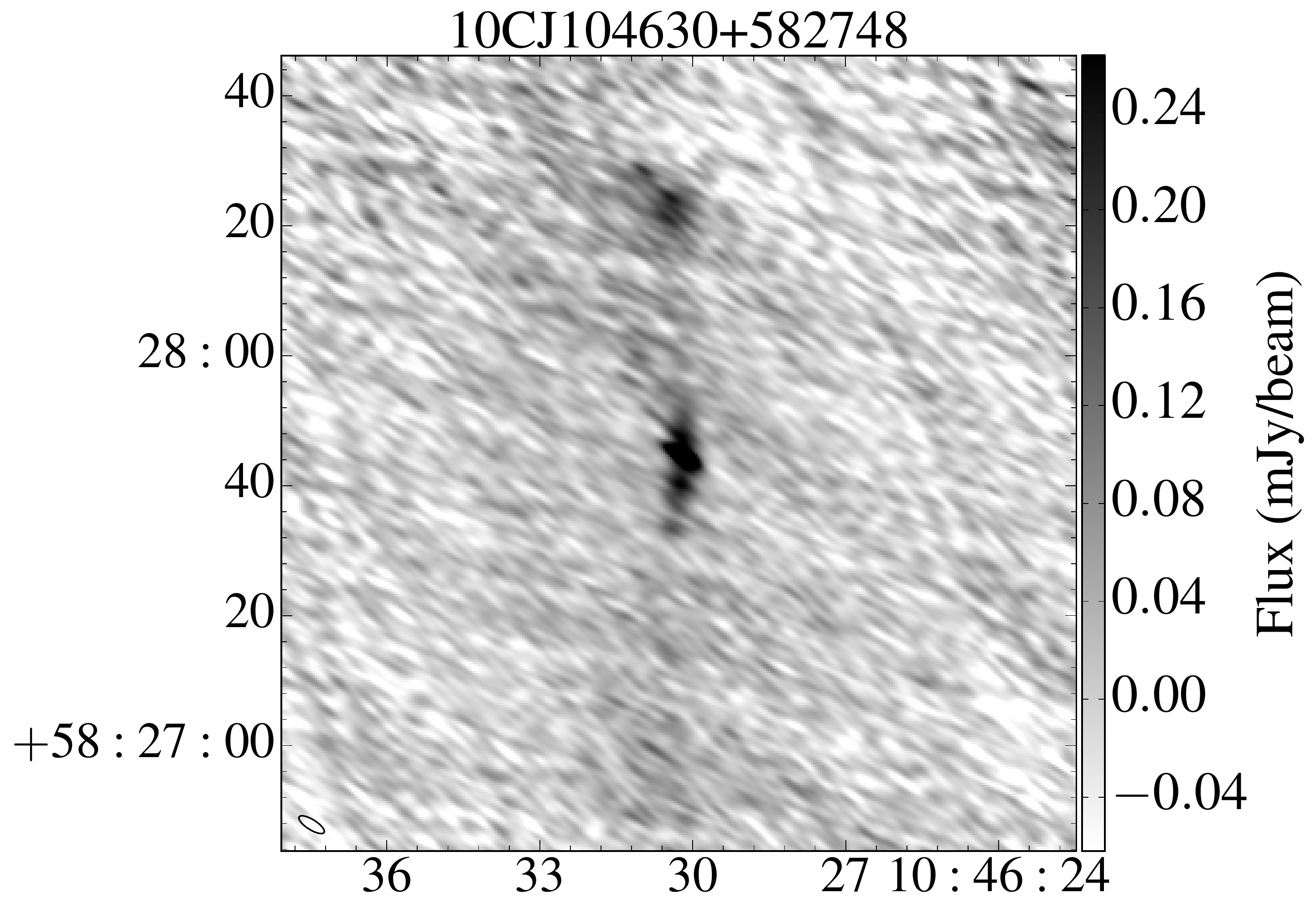}
            \includegraphics[width=6cm]{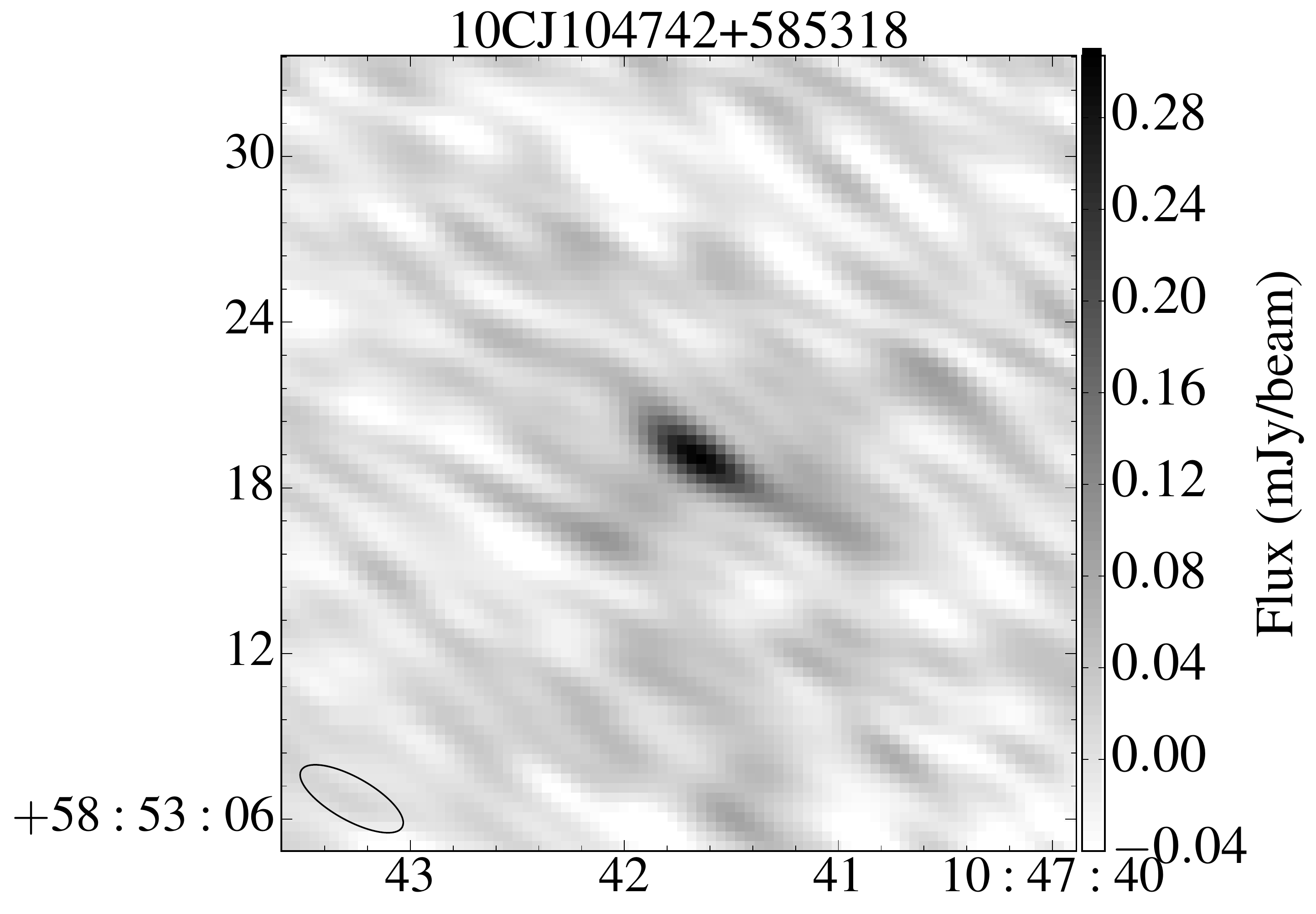}
            \includegraphics[width=6cm]{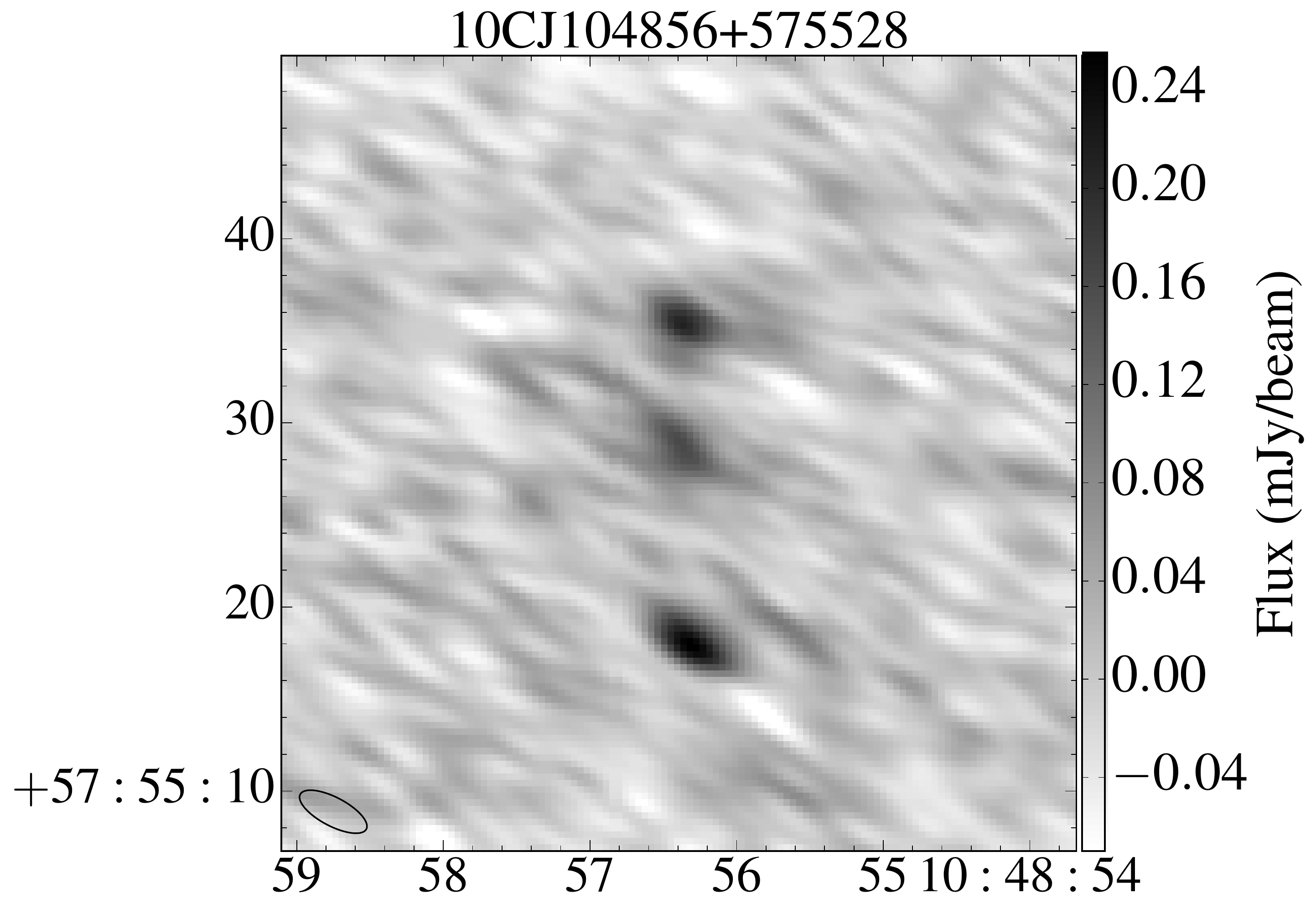}}
\centerline{\includegraphics[width=6cm]{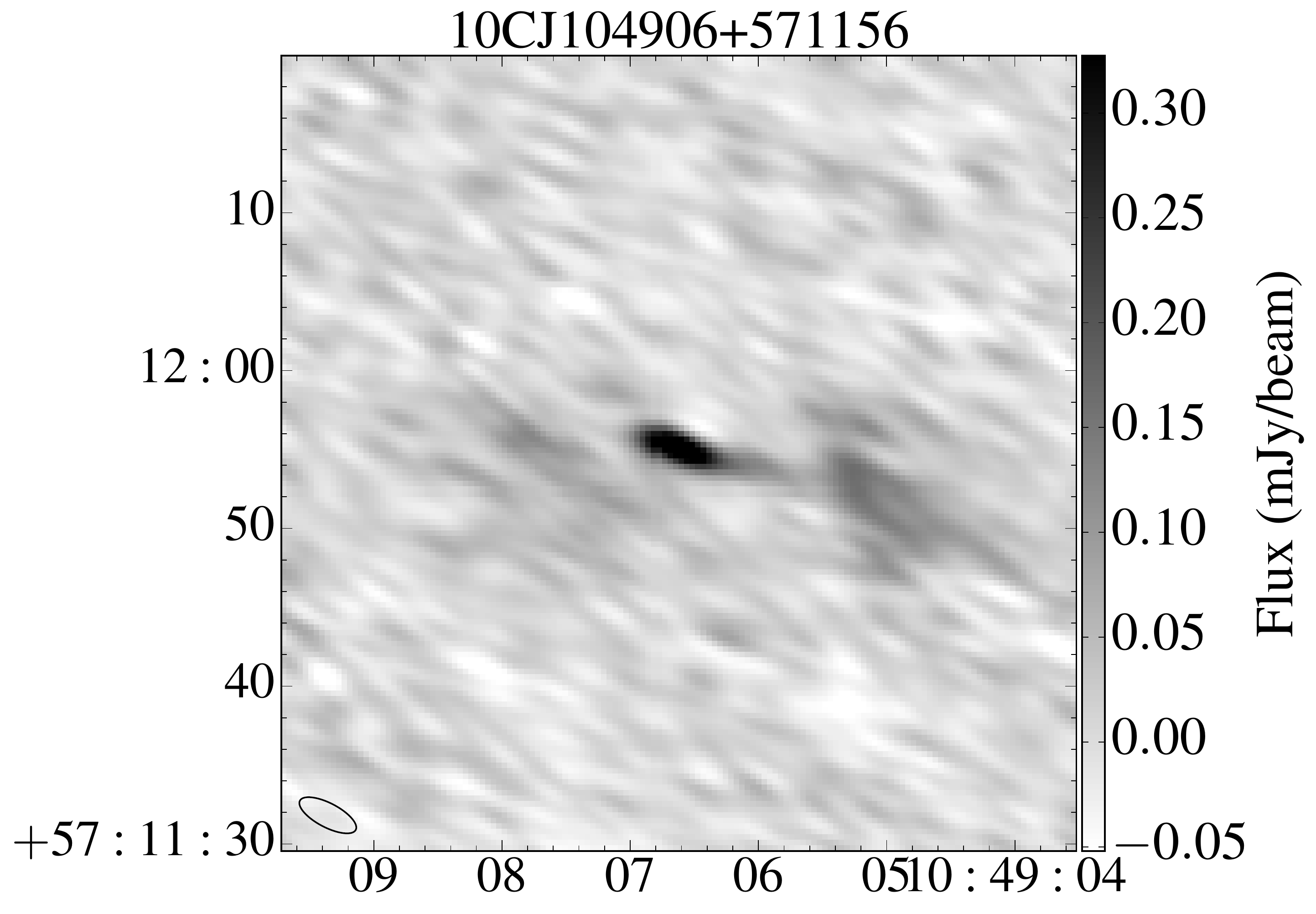}
            \includegraphics[width=6cm]{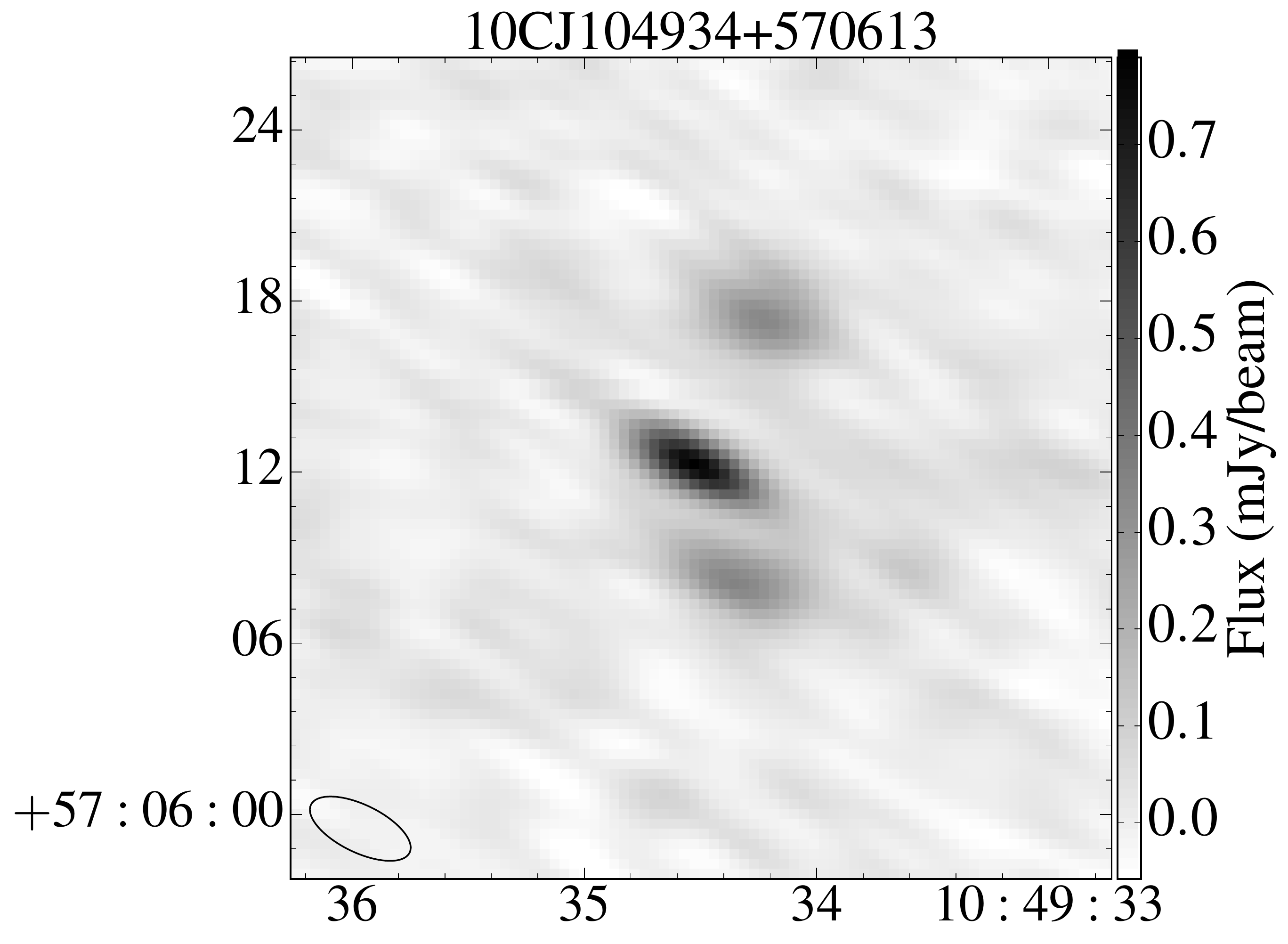}
            \includegraphics[width=6cm]{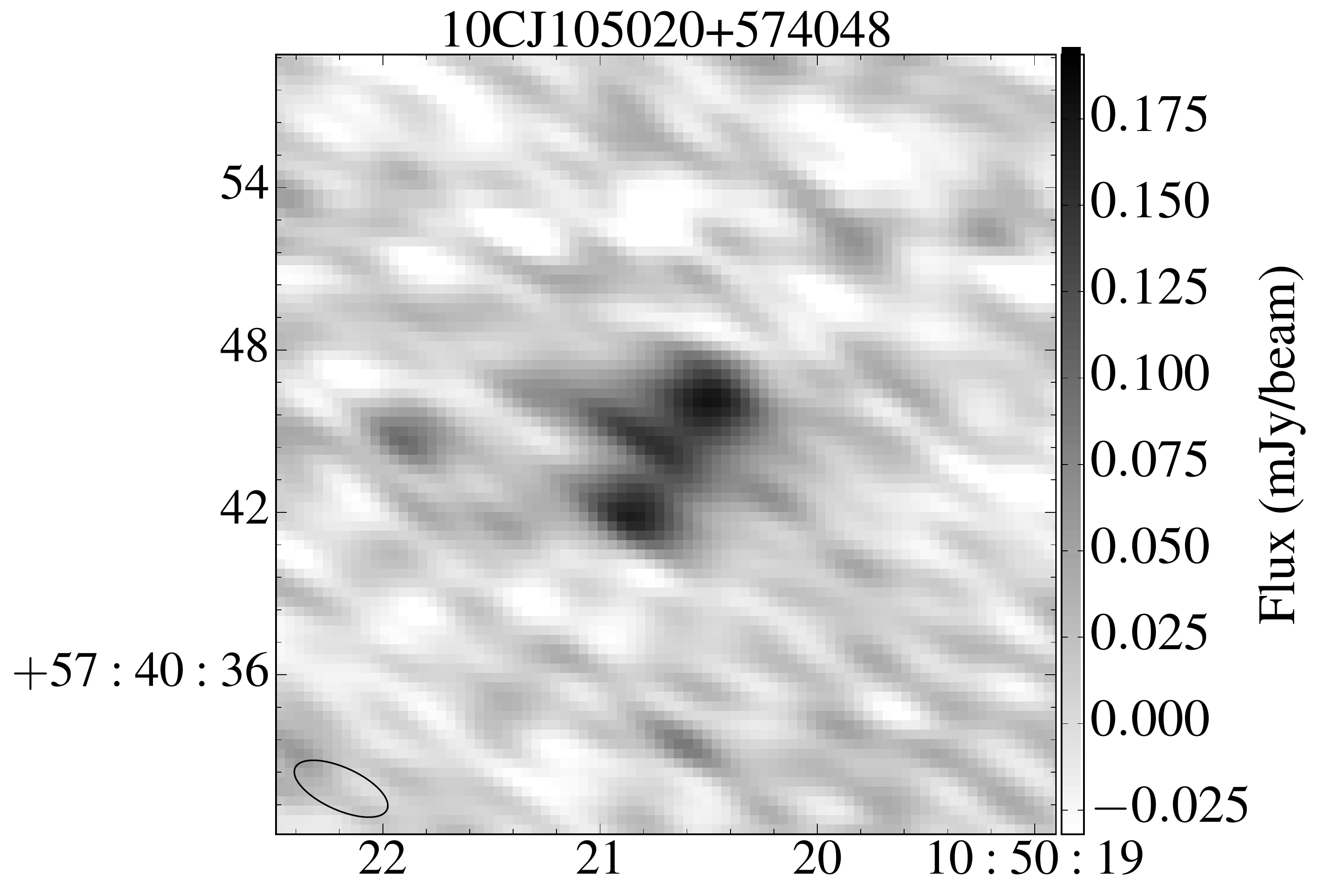}}
\centerline{\includegraphics[width=6cm]{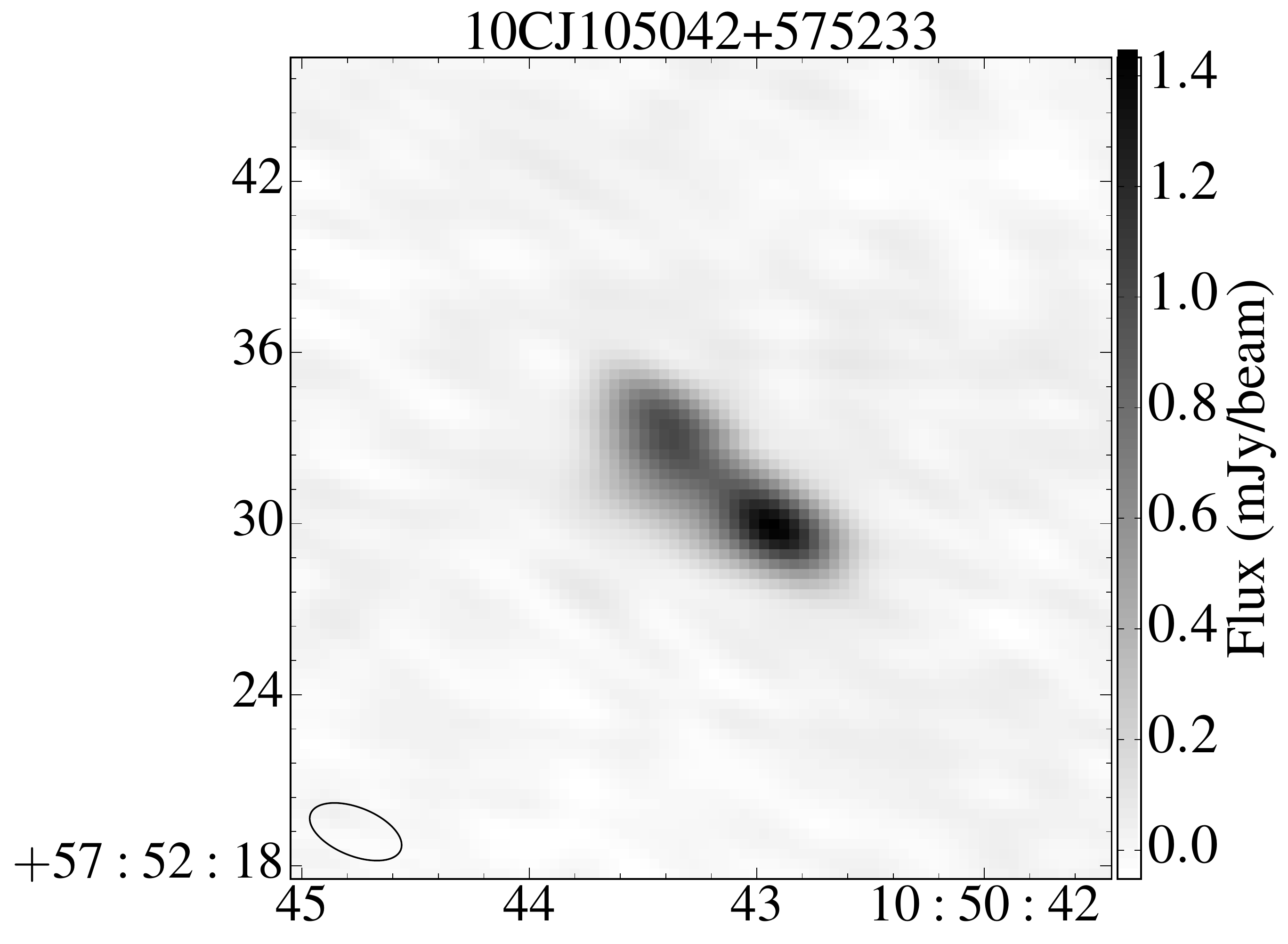}
            \includegraphics[width=6cm]{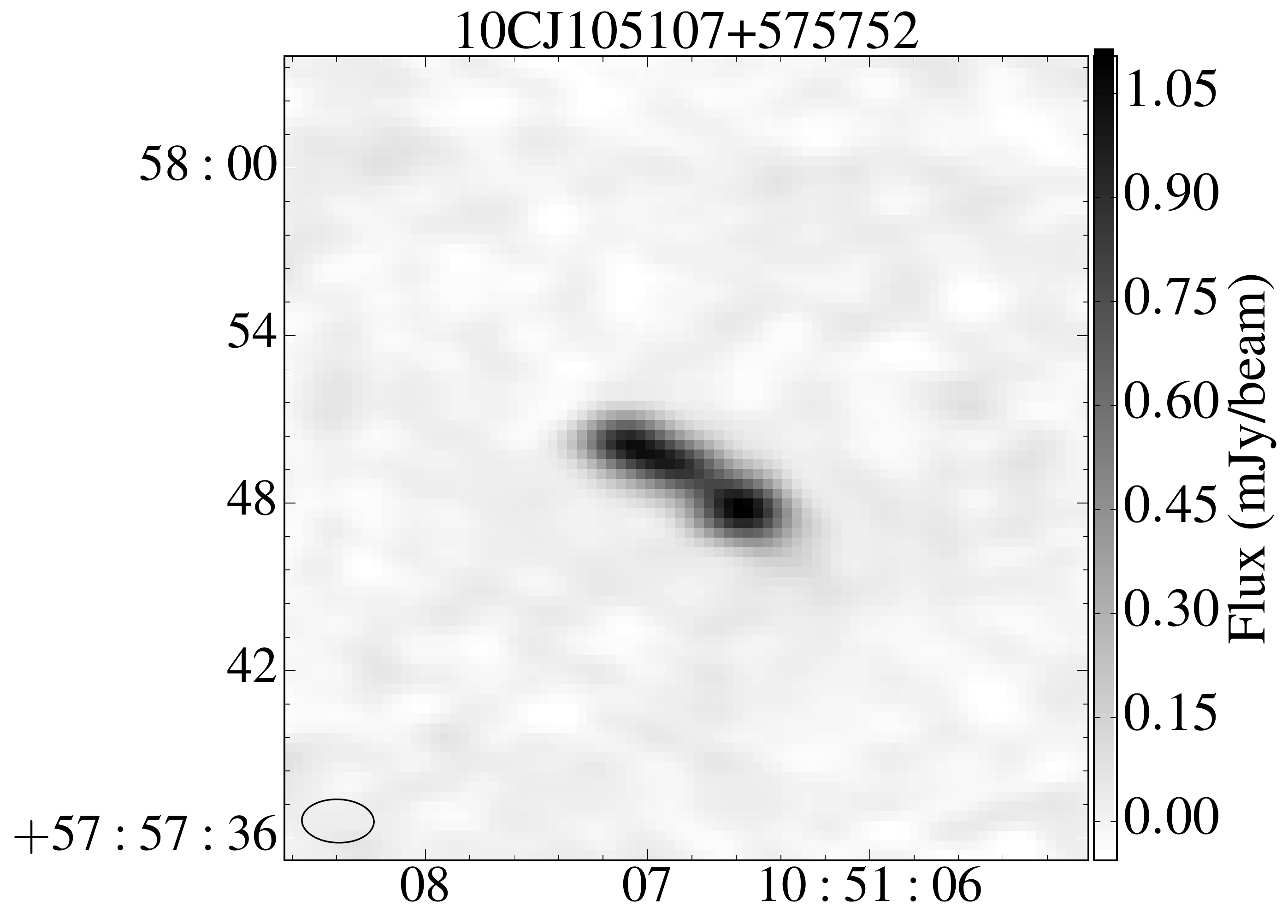}
            \includegraphics[width=6cm]{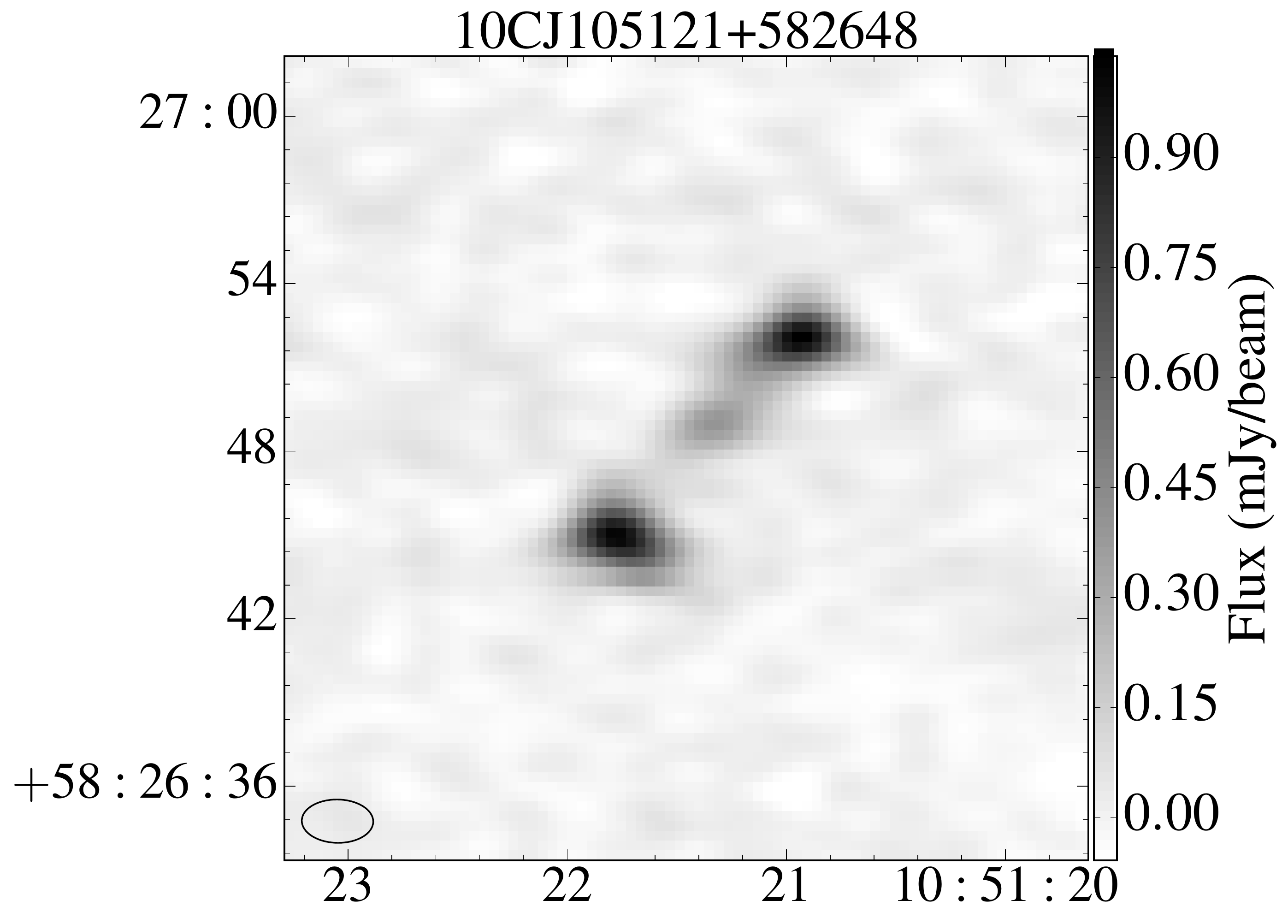}}
\centerline{\includegraphics[width=6cm]{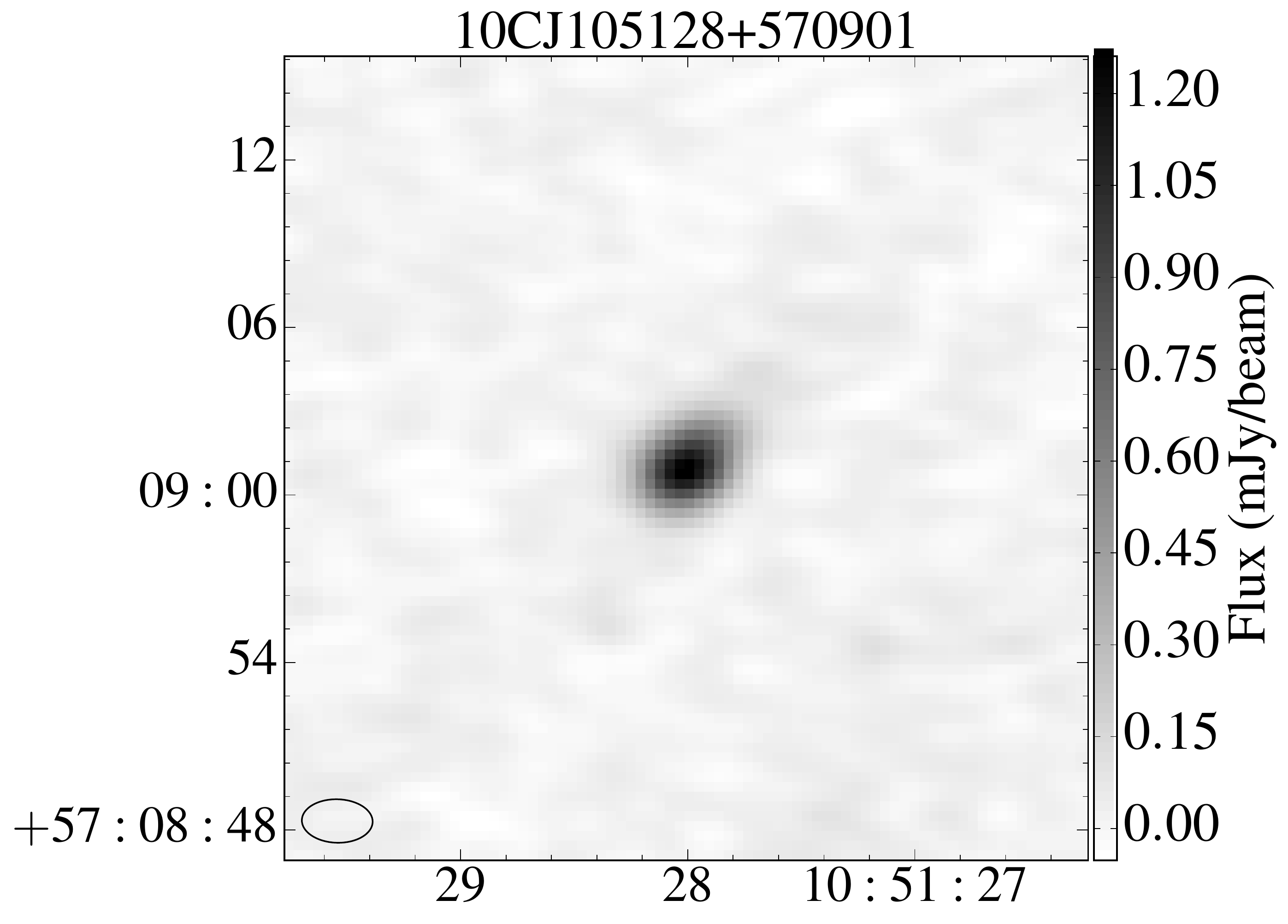}
            \includegraphics[width=6cm]{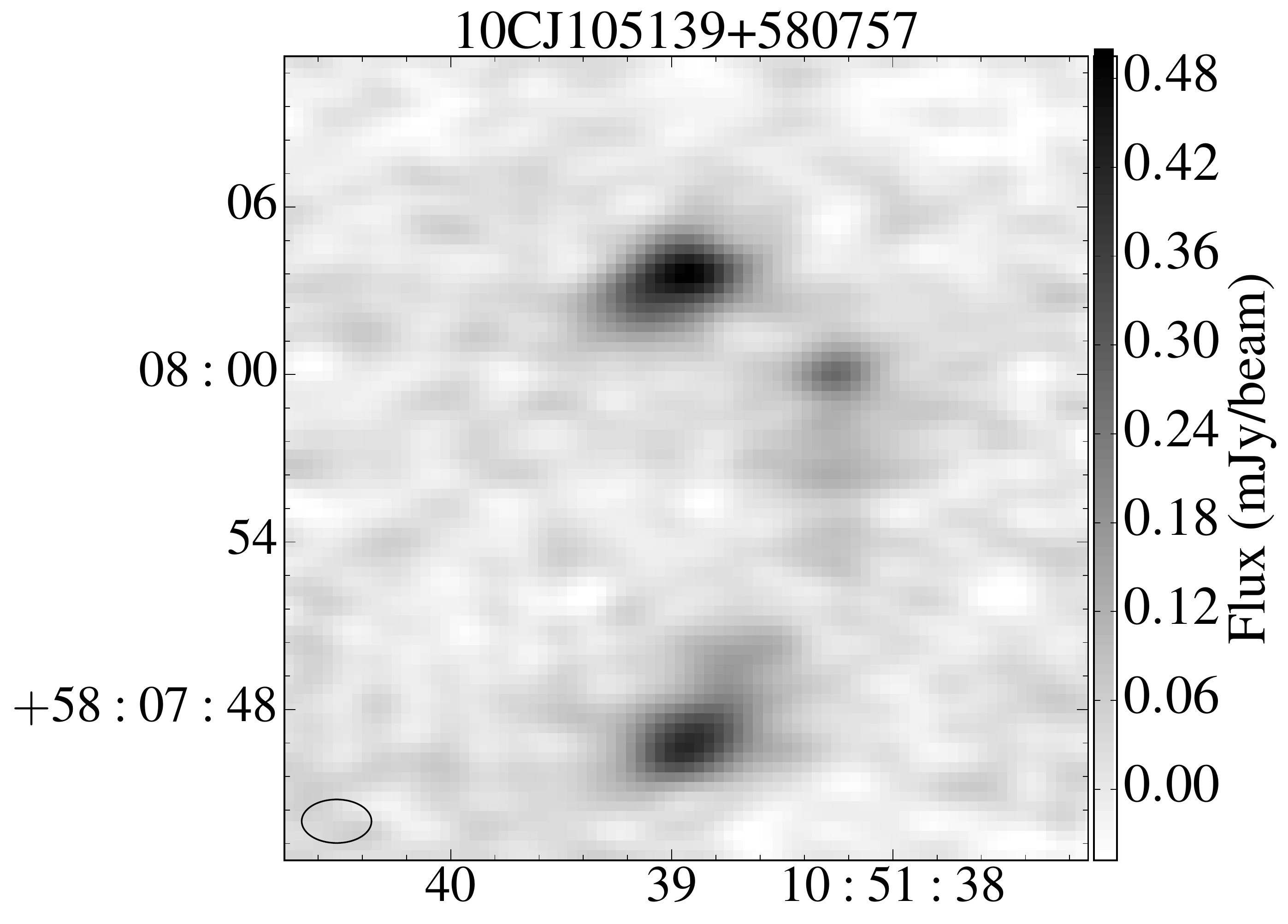}
            \includegraphics[width=6cm]{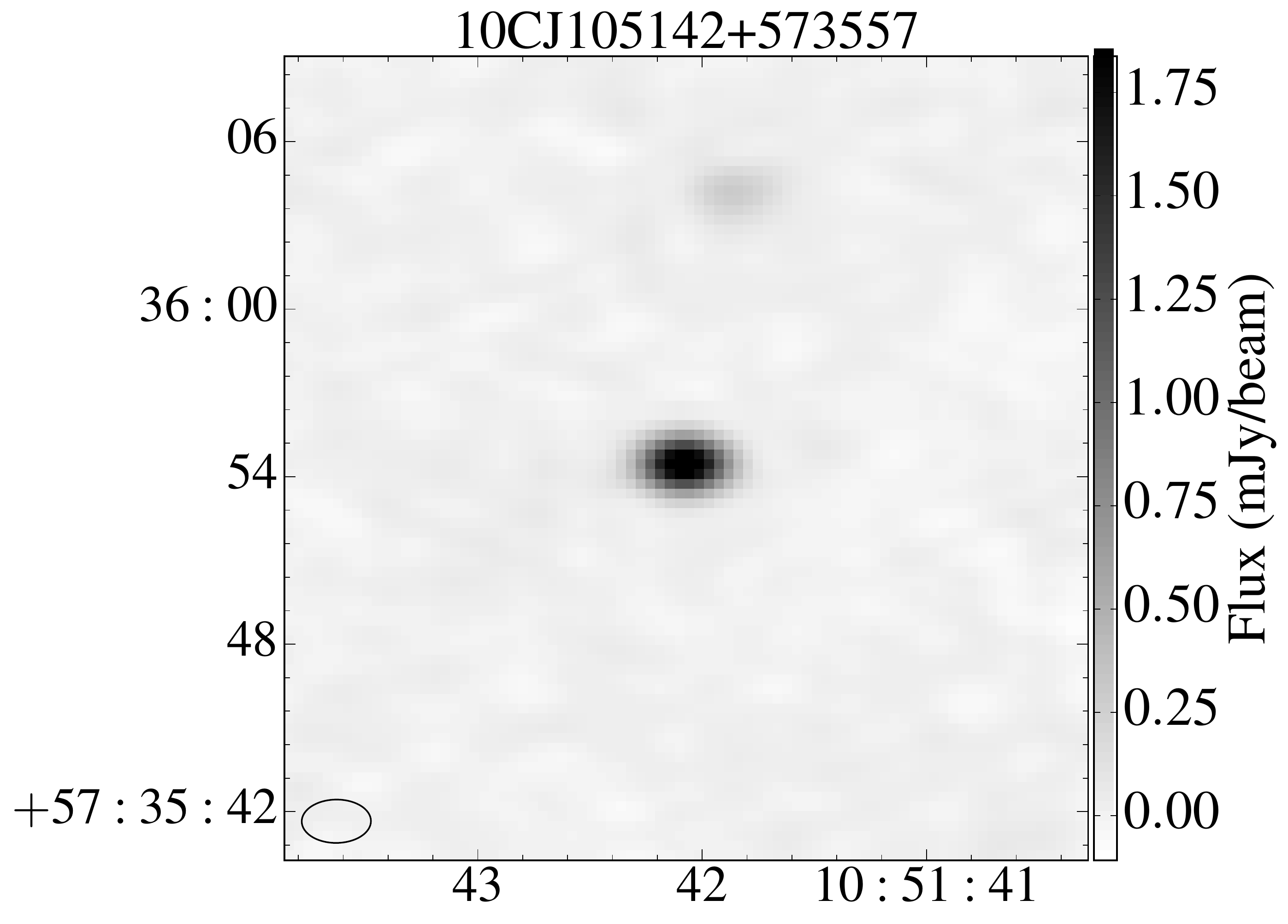}}
\centerline{\includegraphics[width=6cm]{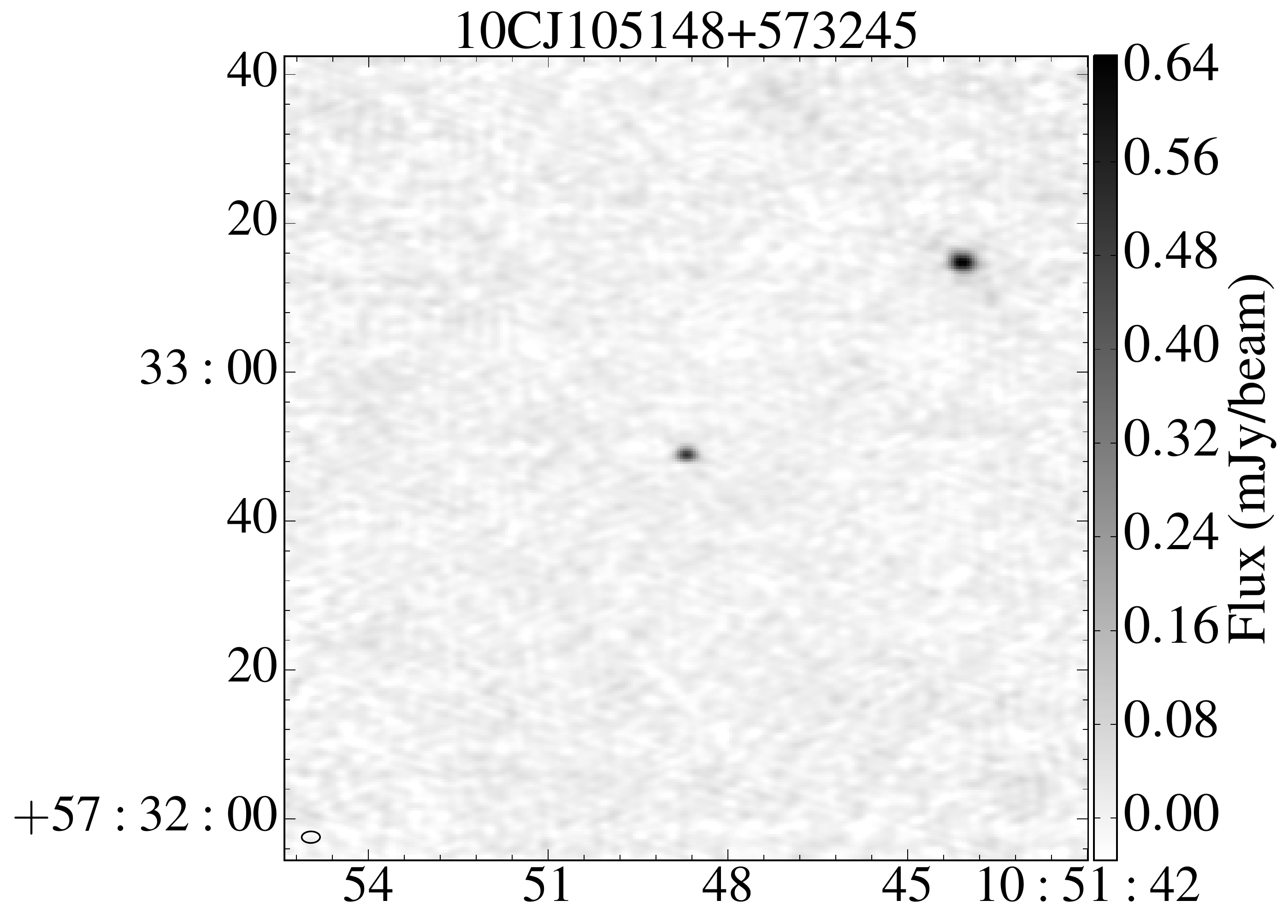}
            \includegraphics[width=6cm]{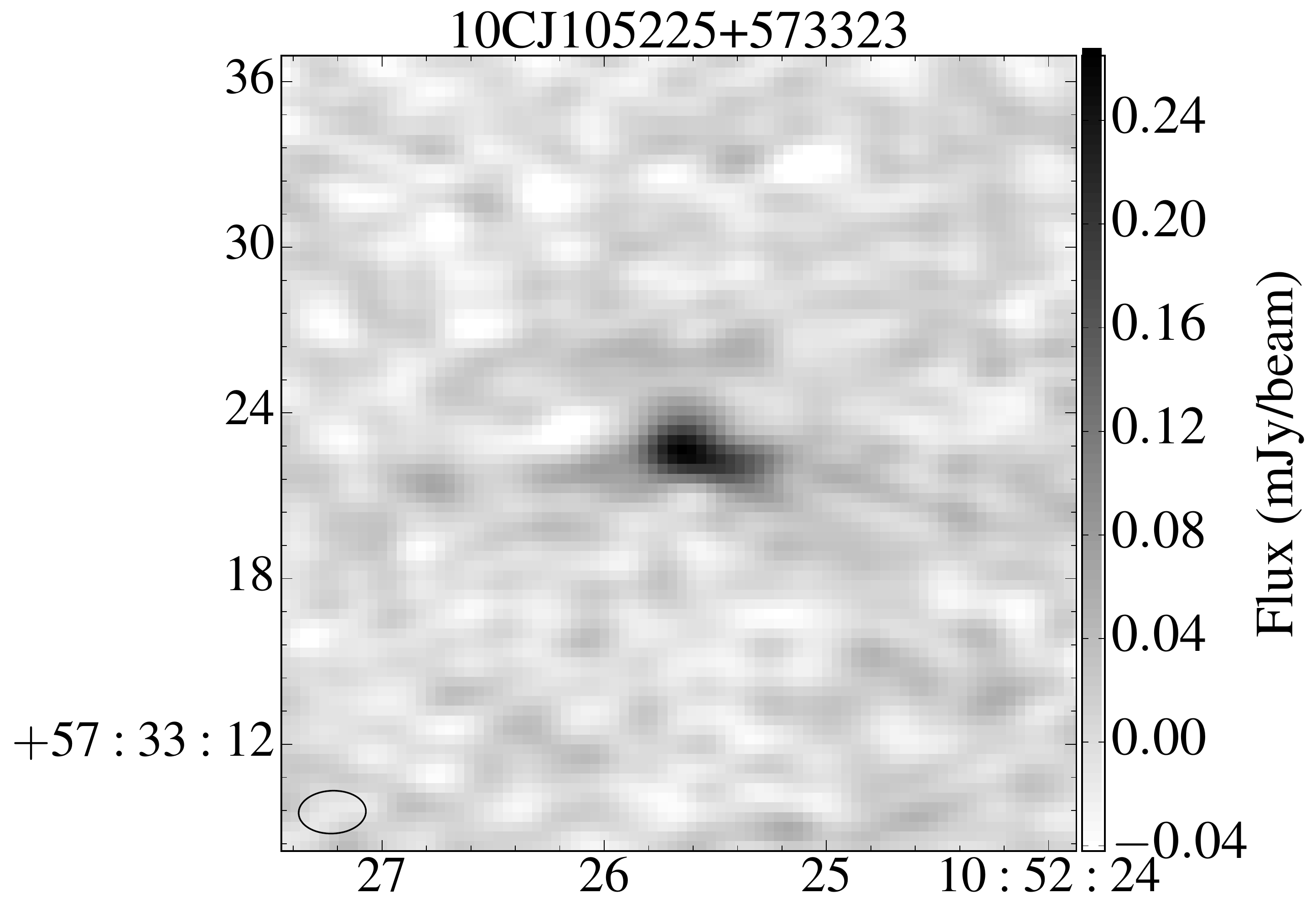}
            \includegraphics[width=6cm]{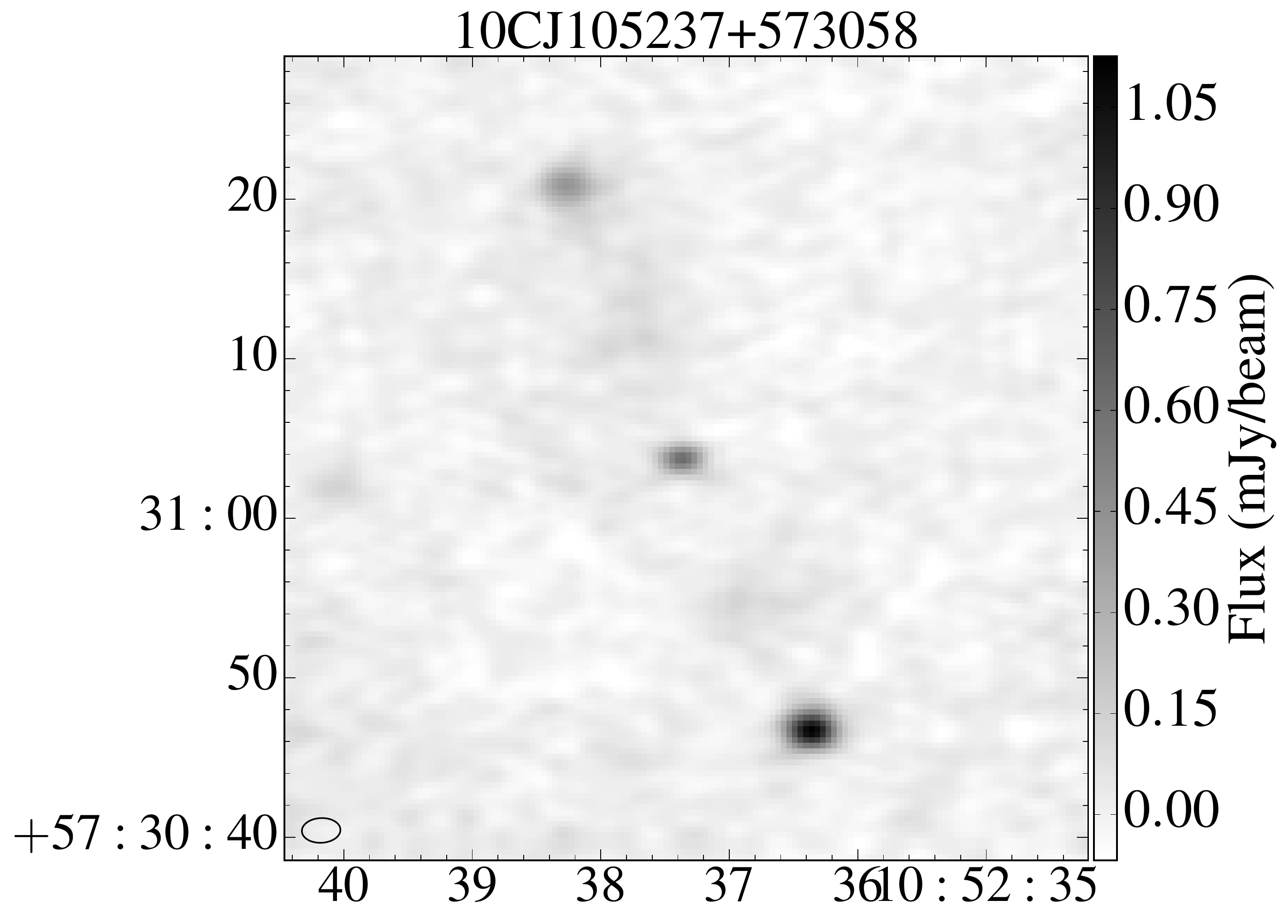}}            
\caption{The new 15~GHz VLA observations of the 22 sources resolved in the VLA observations.}\label{fig:images_extended}
\end{figure*}

\begin{figure*}
\centerline{\includegraphics[width=6cm]{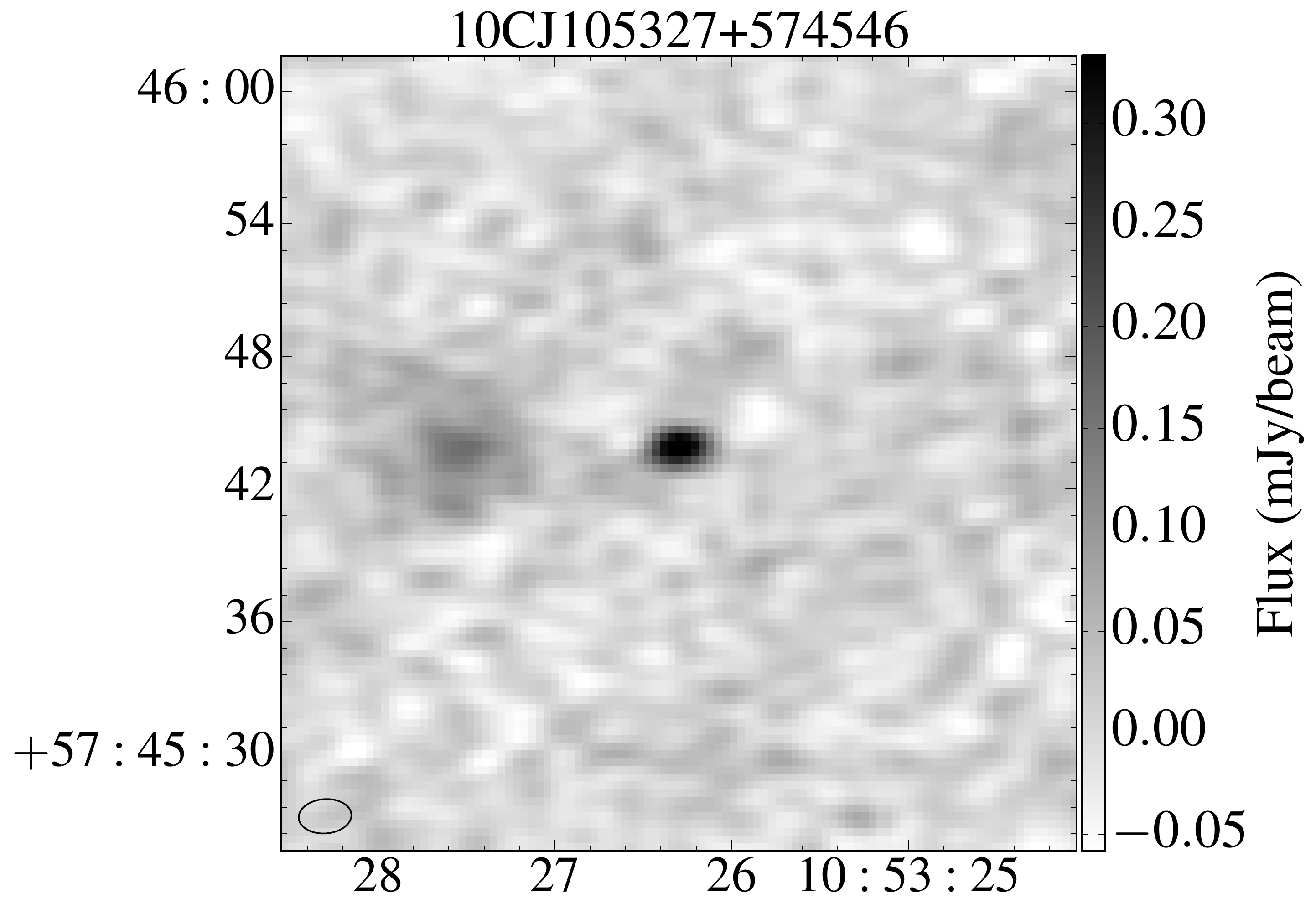}
            \includegraphics[width=6cm]{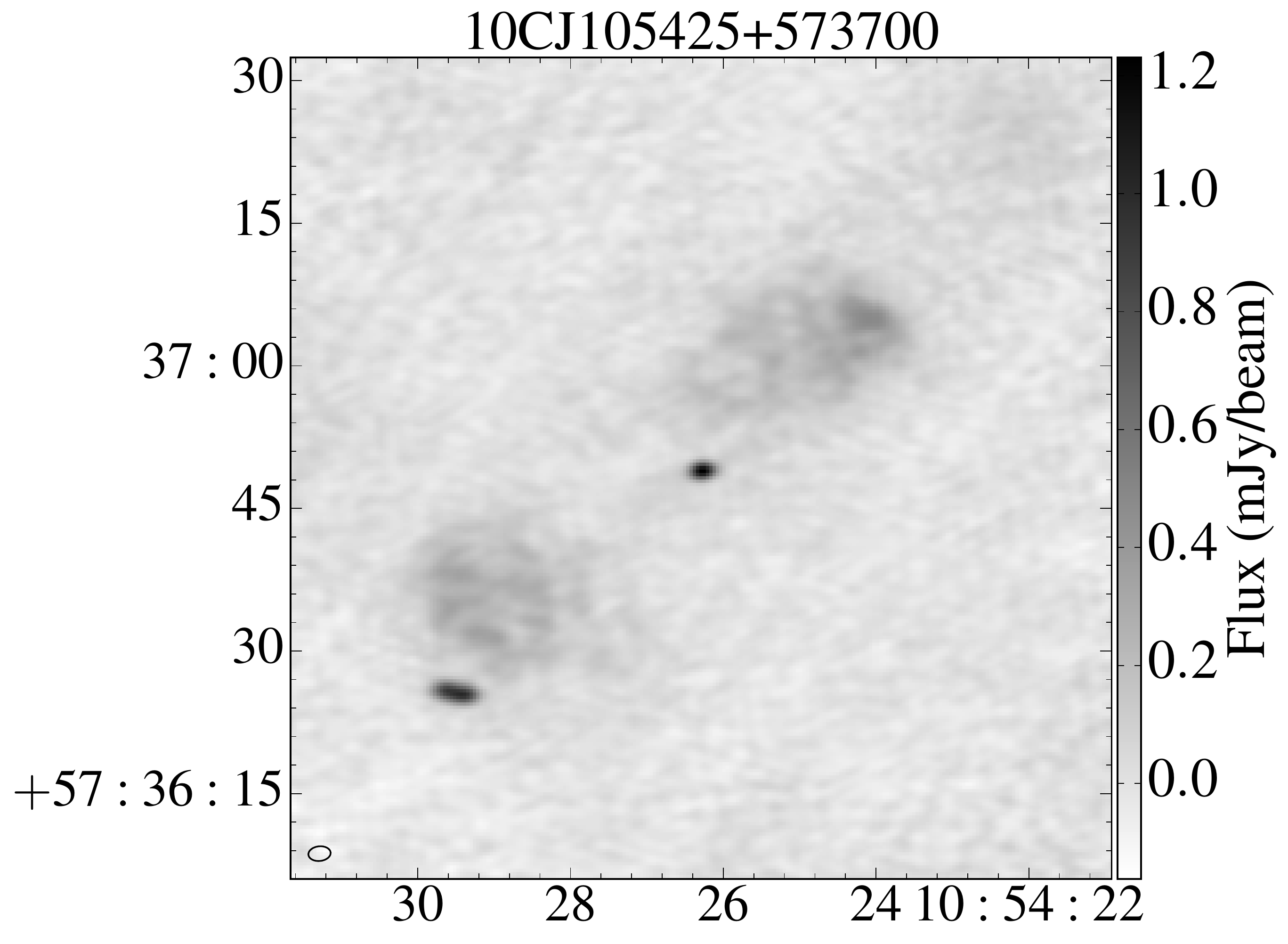}
            \includegraphics[width=6cm]{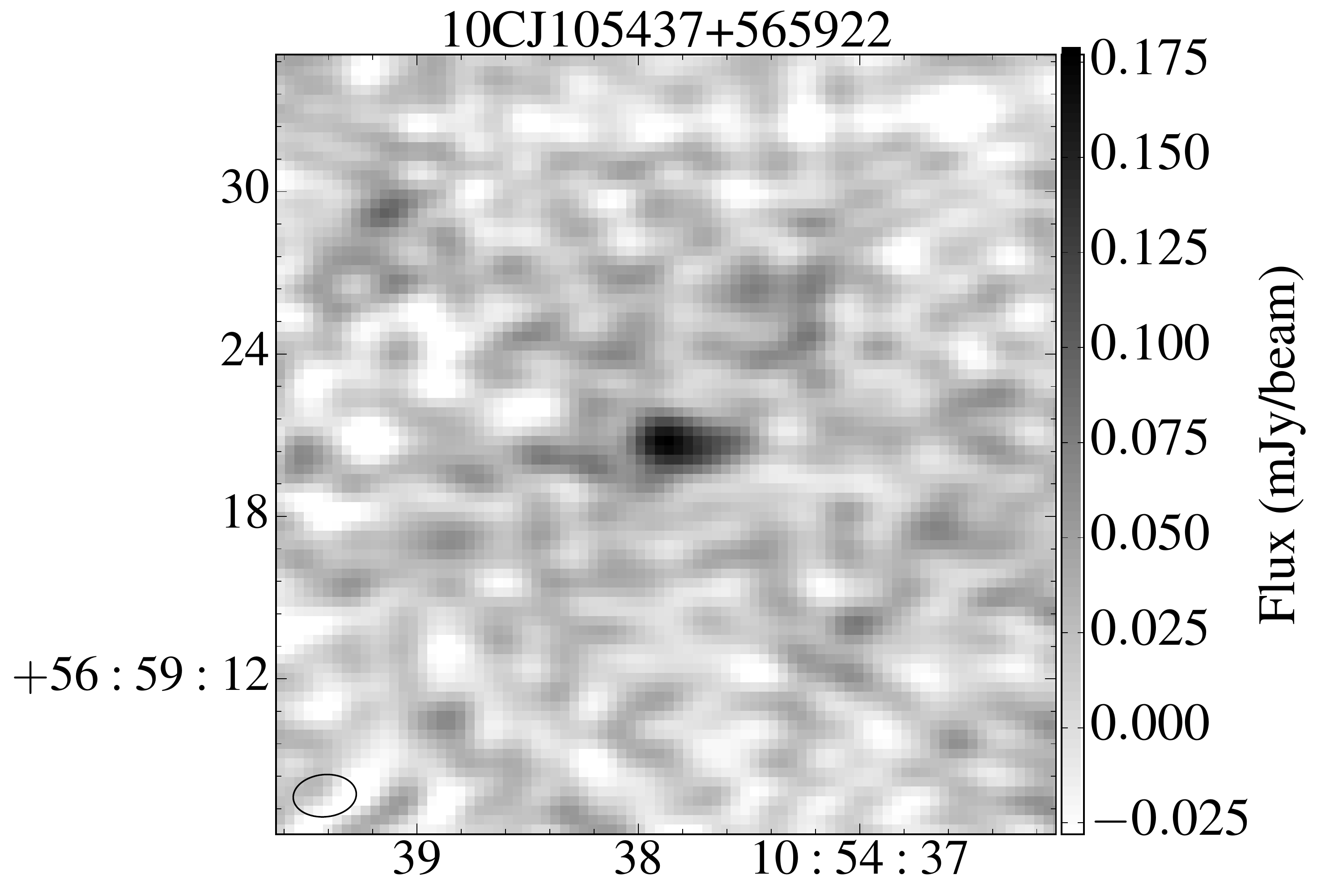}}   
\centerline{\includegraphics[width=6cm]{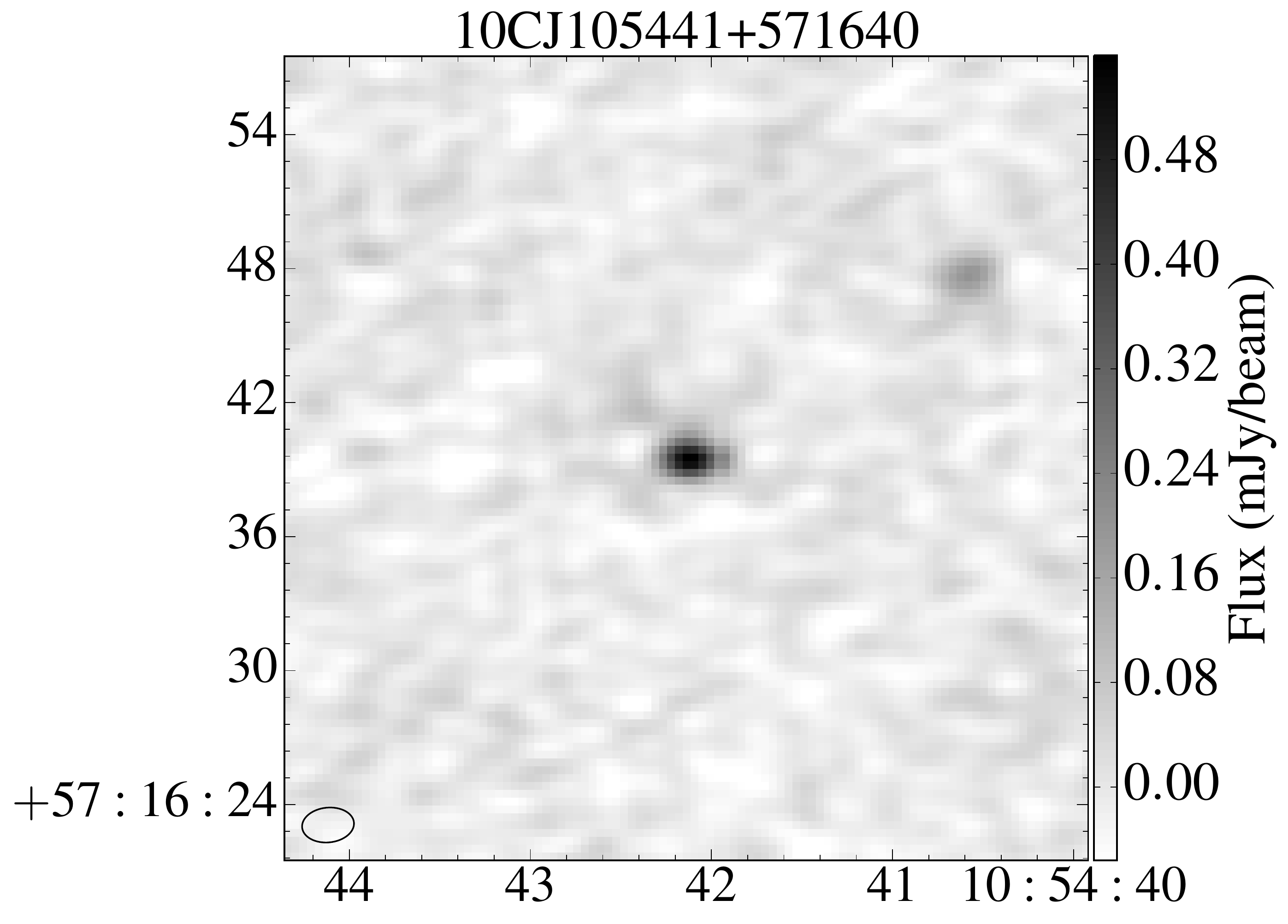}
            \includegraphics[width=6cm]{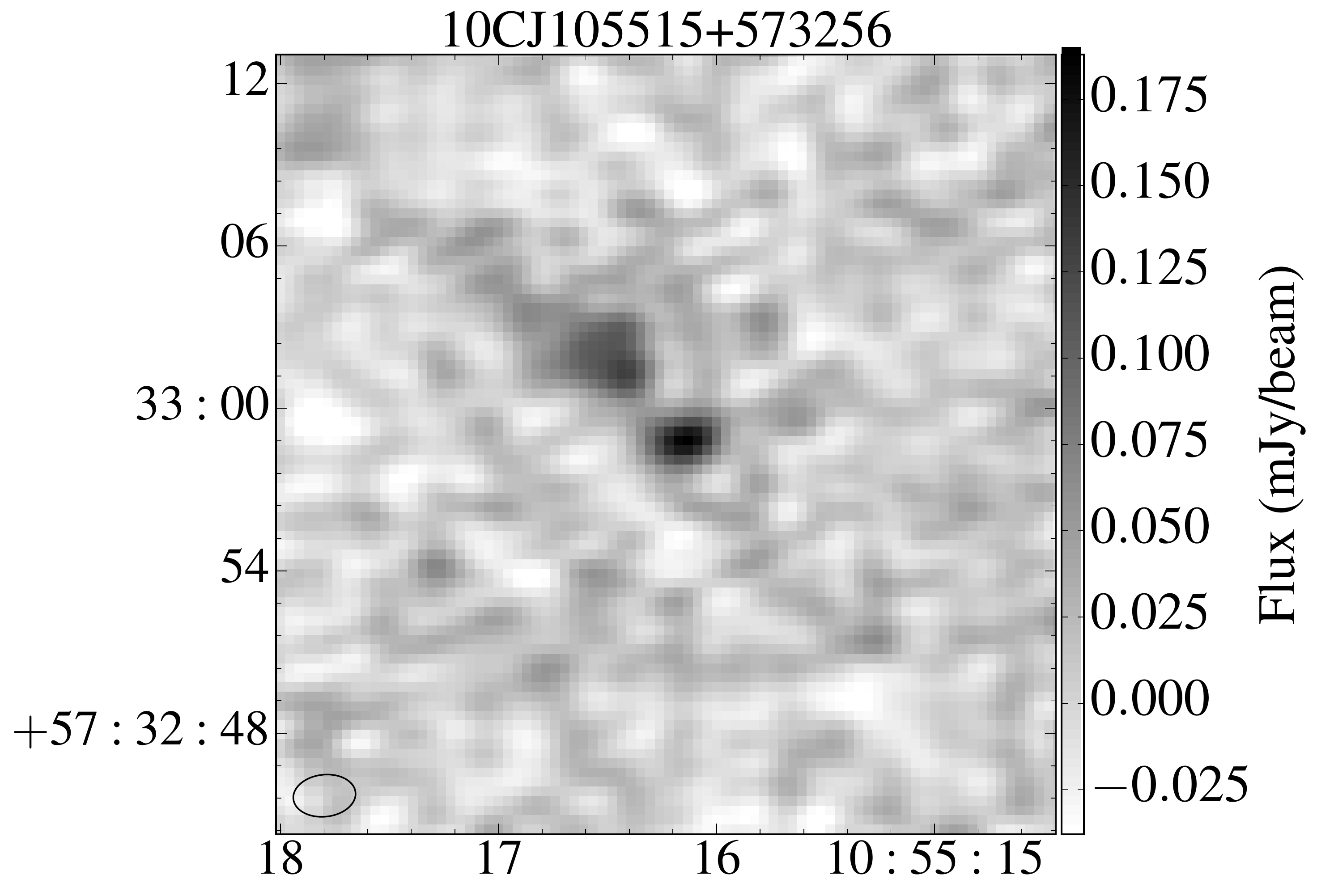}
            \includegraphics[width=6cm]{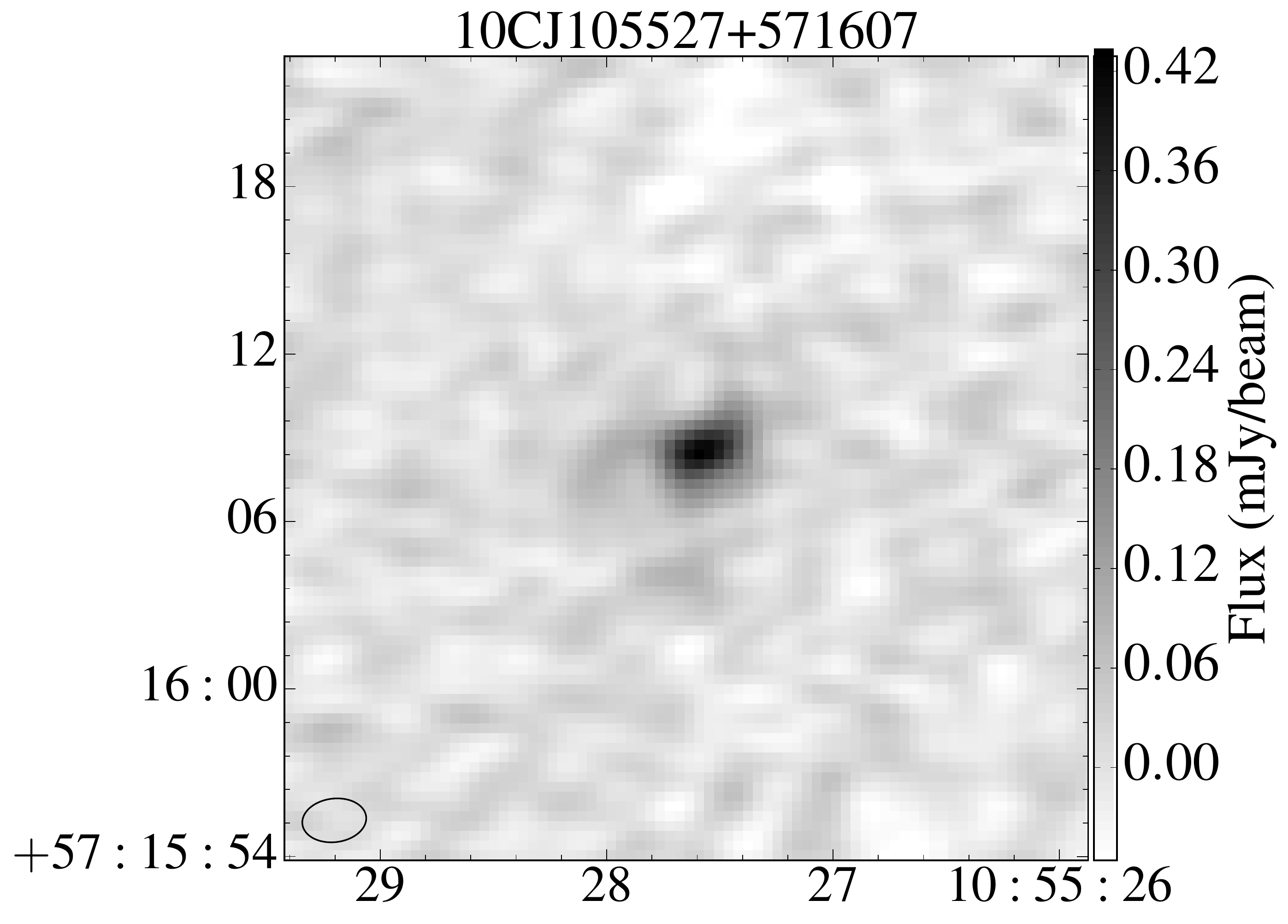}}
\centerline{\includegraphics[width=6cm]{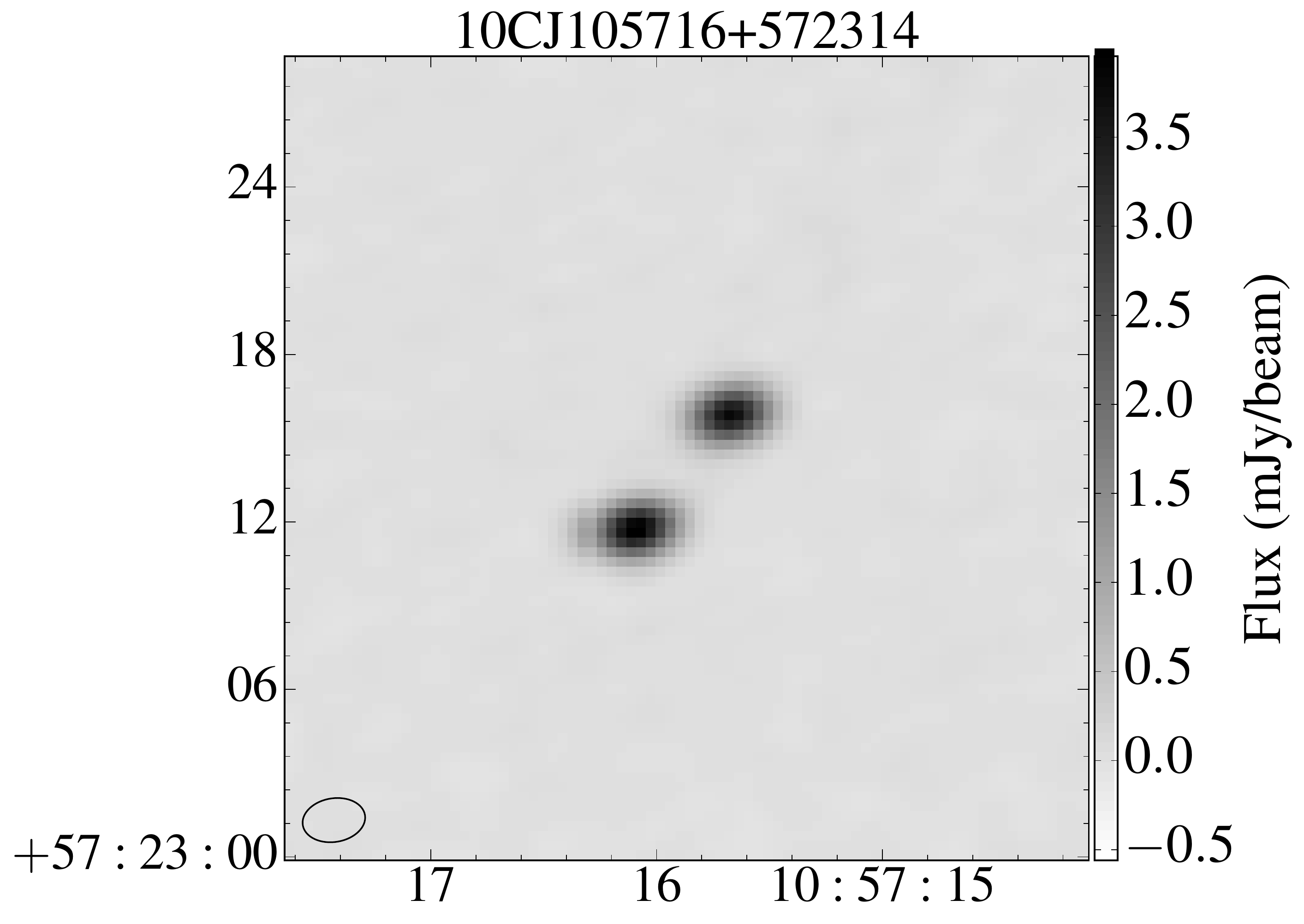}}
\contcaption{}
\end{figure*}

For single component sources angular sizes were estimated using the Gaussian fit from \textsc{imfit}. For those which could be resolved, the deconvolved full-width half maximum (FWHM) major axis was used as a measure of the angular size. If, however, the source could not be deconvolved by \textsc{imfit}, the FWHM major axis of the restoring beam was taken as an upper limit on the angular size of the source. This is a conservative upper limit, as a source must be intrinsically smaller than this to appear unresolved. This was the case for 73 sources (26 in SB 1 and 47 in SB 2). As the synthesised beam size and shape is different for the two SBs ($1.6 \times 2.4$~arcsec$^2$ compared to $1.6 \times 4.0$~arcsec$^2$) there are different upper limits on the angular sizes of the unresolved sources observed in the different SBs. Note that for the second SB the unresolved sources are probably smaller than the 4~arcsec upper limit suggests as the synthesised beam is highly elliptical and the sources are not resolved in either direction.

For the 17 sources resolved into more than one component the angular size was measured from the contour map by taking the maximum straight line extent of the source within the $3\sigma$ contours. The angular sizes of the sample of 95 sources measured from the 15~GHz VLA observations are shown in Fig.~\ref{fig:ang_size} and listed in Table~\ref{tab:sources}.

Six of the sources which are unresolved in the new VLA observations have extended ($>10$~arcsec) emission visible in the lower-frequency observations (at 1.4 GHz and/or 610~MHz, described in Section~\ref{section:sample}) and are therefore clearly not genuinely compact. Therefore 28 sources (22 resolved at 15 GHz and a further 6 resolved at lower frequencies) in this sample are extended and 67 are compact on $\sim$arcsec scales.
Since the extended emission typically has a steep spectrum, the lower-frequency observations provide a better measure of any extended emission (where present) than the 15~GHz observations. Therefore, when considering the overall linear sizes of the sources, the measurements from the lower-frequency surveys (defined in Section 6.4 of \citetalias{2015MNRAS.453.4244W}) are used for the extended sources. For the sources which are compact in both the lower-frequency and 15~GHz observations (67 sources), the 15 GHz measurement is used as an upper limit on the size of the source as this provides a smaller limit. However, note that it is possible that these sources may have some extended emission which is resolved out of the new VLA observations, this is discussed further in Section~\ref{section:10C}.

In \citetalias{2016MNRAS.462.2122W} the sources in this sample were classified according to their morphology and radio spectra using 610~MHz GMRT data with a resolution of $5 \times 6$ arcsec$^2$. Each source was classified as FR0, FR0/I, FRI, FRI/II or FRII based on its morphology; the compact FR0 sources were further classified based on their radio spectra and linear sizes as either FR0c (candidate compact steep spectrum (CSS) sources), FR0g (candidate GHz-peaked spectrum (GPS) sources), or FR0u (uncertain) (detailed descriptions of the classification criteria are given in \citetalias{2016MNRAS.462.2122W}). These classifications were reviewed in light of the new data by examining each source by eye. Nine sources which were previously classified as FR0u are resolved by the new observations and therefore re-classified. A further three sources with uncertain classifications (FR0/I or FRI/II) are also re-classified. Note that none of the sources previously identified as candidate CSS or GPS sources are resolved in the new observations, providing support for these classifications.
For the six sources which appear compact in the new observations where only the radio core is detected but which have extended emission visible at 610~MHz the original 610~MHz FRI or II classifications are retained.

When examining all available images we realised that two nearby sources (10CJ105144+573313 and 10CJ105148+573245, see Fig.~\ref{fig:images_extended}) which had previously been classified as two FR0 sources are more likely the core and hotspot of one source. We have therefore combined these sources into one entry in the catalogue, labelled as 10CJ105148+573245 as this source is the core.
The FR classification has also changed for one other source, 10CJ105009+570724, which was previously thought to be one FRI source but is now classified as two separate FR0 sources. This is discussed further in Section~\ref{section:opt-matching}. These two cases illustrate the difficulties in correctly classifying and identifying host galaxies for radio sources.

The properties of the sources in this sample derived from the new VLA observations are summarised in Table~\ref{tab:sources}, which is available in the online supplementary material in machine-readable format. Images of the 22 sources which are resolved in the observations are shown in Fig.~\ref{fig:images_extended}.

\subsection{False detection}

One source in the sample, 10CJ104927+583830, was not detected in the new VLA observations. This source had a peak flux density of $0.45 \pm 0.08$~mJy/beam in the 10C catalogue but is undetected in the new image which has an rms noise of 0.018~mJy, giving a $5\sigma$ upper limit on the peak flux density of the source of 0.09~mJy/beam. This suggests that if the source is real, its peak flux density must have decreased by more than a factor of five over a $\sim9$~year period. 

However, we now believe that this source is a false detection. It is the only source in the sample which was not detected in any of the other lower-frequency radio catalogues which cover this field, and it also does not have a counterpart in the SERVS Data Fusion multi-wavelength catalogue \citepalias{2015MNRAS.453.4244W}. When extracting sources for the 10C catalogue, exclusion zones were defined around bright ($>15$~mJy) sources due to the increased noise close to these sources caused by amplitude, phase and deconvolution errors (see \citealt{2011MNRAS.415.2708D} for details). This source is close to a bright source and fell just outside one such exclusion zone; the exclusion zone in question had a radius of 3.35~arcmin and this source is located 3.42~arcmin from centre of the zone. On examining the original 10C map, the noise level appears to still be elevated in the region around this source. Thus all the evidence indicates that this source is a false detection, and it is therefore removed from the sample discussed in the remainder of this paper.

\begin{table*}
\caption{Properties of the sources in this sample based on the new VLA observations. A machine-readable version of this table is available online as supplementary material.}\label{tab:sources}
\small\rm
\medskip
\centering
\renewcommand{\tabcolsep}{1.3mm}
\begin{tabular}{cccddddcdddc}\hline
10C ID & RA & Dec & \dhead{$S_\textrm{peak}$} & \dhead{$\Deltaup S_\textrm{peak}$} & \dhead{$S_\textrm{int}$} & \dhead{$\Deltaup S_\textrm{int}$} & \dhead{$N_\textrm{comp}$} & \dhead{rms / $\muup$Jy} & \dhead{$S_\textrm{core}$} &  \dhead{$\theta$} & FR\\
       &    &     & \dhead{/ mJy}             & \dhead{/ mJy} 					  & \dhead{/ mJy} & \dhead{/ mJy} & & \dhead{beam$^{-1}$} & \dhead{/ mJy} & \dhead{/ arcsec} & class\\
(1) & (2) & (3) & \dhead{(4)} & \dhead{(5)} & \dhead{(6)} & \dhead{(7)} & (8) & (9) & \dhead{(10)} & \dhead{(11)} & (12)\\\hline
10C J104320+585621 & 10:43:20.515 & +58:56:20.780 & 1.043 & 0.022 & 1.052 & 0.046 & 1 & 34 &    & < 4.6 & FRII \\
10C J104328+590312 & 10:43:28.435 & +59:03:14.080 & 1.193 & 0.020 & 1.262 & 0.044 & 1 & 37 &    & < 4.0 & FR0u \\
10C J104344+591503 & 10:43:44.677 & +59:15:02.550 & 2.686 & 0.020 & 2.724 & 0.042 & 1 & 27 &    & < 4.7 & FR0u \\
10C J104428+591540 & 10:44:28.382 & +59:15:41.490 & 0.512 & 0.020 & 0.571 & 0.045 & 1 & 35 &    & < 5.1 & FR0g \\
10C J104441+591949 & 10:44:41.132 & +59:19:46.100 & 0.643 & 0.019 & 0.639 & 0.038 & 1 & 26 &    & < 4.5 & FR0g \\
10C J104451+591929 & 10:44:50.388 & +59:19:26.350 & 1.116 & 0.023 & 1.533 & 0.055 & 1 & 29 &    & < 4.0 & FR0u \\
10C J104528+591328 & 10:45:28.206 & +59:13:26.190 & 1.768 & 0.022 & 1.800 & 0.045 & 1 & 27 &    & < 4.5 & FR0c \\
10C J104539+585730 & 10:45:39.678 & +58:57:29.550 & 0.417 & 0.046 & 0.731 & 0.203 & 1 & 32 &    & < 5.1 & FRII \\
10C J104551+590838 & 10:45:50.787 & +59:08:41.420 & 0.695 & 0.016 & 0.757 & 0.036 & 1 & 23 &    & < 4.0 & FR0u \\
10C J104624+590447 & 10:46:24.841 & +59:04:45.760 & 0.896 & 0.025 & 1.214 & 0.059 & 1 & 23 &    & < 4.0 & FR0u \\
10C J104630+582748 & 10:46:30.253 & +58:27:40.360 & 0.263 & 0.037 & 12.260 & 2.332 & 5 & 31 & 1.079 & 110.0 & FRII \\
10C J104633+585816 & 10:46:33.140 & +58:58:15.160 & 0.272 & 0.019 & 0.250 & 0.039 & 1 & 29 &    & < 4.5 & FR0g \\
10C J104648+590956 & 10:46:49.522 & +59:09:55.840 & 0.824 & 0.020 & 0.811 & 0.042 & 1 & 30 &    & < 4.6 & FR0u \\
10C J104700+591903 & 10:47:00.681 & +59:19:01.240 & 2.473 & 0.024 & 2.635 & 0.050 & 1 & 27 &    & < 4.0 & FR0c \\
10C J104710+582821 & 10:47:10.910 & +58:28:21.190 & 18.898 & 0.050 & 19.641 & 0.103 & 1 & 34 &    & < 4.0 & FR0u \\
10C J104718+585119 & 10:47:18.620 & +58:51:18.770 & 0.947 & 0.023 & 0.963 & 0.047 & 1 & 28 &    & < 4.4 & FRI \\
10C J104719+582114 & 10:47:19.250 & +58:21:14.130 & 72.649 & 0.302 & 81.427 & 0.632 & 1 & 138 &    & < 4.4 & FR0u \\
10C J104733+591244 & 10:47:34.481 & +59:12:40.720 & 0.495 & 0.020 & 0.520 & 0.043 & 1 & 27 &    & < 4.0 & FR0u \\
10C J104737+592028 & 10:47:38.521 & +59:20:24.380 & 0.285 & 0.015 & 0.316 & 0.034 & 1 & 18 &    & < 4.0 & FR0c \\
10C J104741+584811 & 10:47:41.660 & +58:48:11.930 & 0.495 & 0.022 & 0.643 & 0.050 & 1 & 23 &    & < 4.0 & FR0u \\
10C J104742+585318 & 10:47:41.659 & +58:53:18.810 & 0.307 & 0.023 & 0.530 & 0.080 & 1 & 26 &    & 5.9 & FRII \\
10C J104751+574259 & 10:47:51.377 & +57:42:54.560 & 1.351 & 0.024 & 1.480 & 0.050 & 1 & 26 &    & < 4.0 & FR0u \\
10C J104802+574117 & 10:48:02.344 & +57:41:25.400 & 0.801 & 0.023 & 0.858 & 0.048 & 1 & 24 &    & < 4.0 & FR0u \\
10C J104822+582436 & 10:48:22.452 & +58:24:34.810 & 2.090 & 0.027 & 2.603 & 0.062 & 1 & 29 &    & < 4.0 & FR0c \\
10C J104824+583029 & 10:48:23.896 & +58:30:27.170 & 2.907 & 0.03 & 3.528 & 0.068 & 1 & 27 &    & < 4.0 & FR0u \\
10C J104826+584838 & 10:48:26.390 & +58:48:41.300 & 0.148 & 0.034 & 0.208 & 0.108 & 1 & 20 &    & < 5.4 & FR0u \\
10C J104836+591846 & 10:48:36.986 & +59:18:46.200 & 0.709 & 0.016 & 0.717 & 0.032 & 1 & 20 &    & < 4.0 & FR0u \\
10C J104844+582309 & 10:48:44.000 & +58:23:06.640 & 2.358 & 0.032 & 2.876 & 0.072 & 1 & 27 &    & < 4.0 & FR0u \\
10C J104849+571417 & 10:48:49.815 & +57:14:15.380 & 1.533 & 0.048 & 2.154 & 0.143 & 1 & 25 &    & < 4.0 & FR0u \\
10C J104856+575528 & 10:48:56.348 & +57:55:17.610 & 0.254 & 0.048 & 0.896 & 0.298 & 3 & 35 & 0.265 & 20.0 & FRI \\
10C J104857+584103 & 10:48:58.050 & +58:41:03.030 & 1.265 & 0.023 & 1.600 & 0.050 & 1 & 26 &    & < 4.0 & FR0c \\
10C J104906+571156 & 10:49:06.641 & +57:11:54.600 & 0.491 & 0.025 & 2.762 & 0.590 & 3 & 26 & 0.510 & 32.0 & FRII \\
10C J104918+582801 & 10:49:18.995 & +58:28:00.790 & 7.494 & 0.035 & 7.809 & 0.069 & 1 & 30 &    & < 4.0 & FR0c \\
10C J104934+570613 & 10:49:34.549 & +57:06:12.290 & 0.790 & 0.044 & 1.756 & 0.172 & 3 & 29 & 0.767 & 13.0 & FRII \\
10C J104939+583530 & 10:49:39.970 & +58:35:30.530 & 20.368 & 0.084 & 24.545 & 0.178 & 1 & 48 &    & < 4.0 & FR0c \\
10C J104943+571739 & 10:49:43.777 & +57:17:37.040 & 1.751 & 0.020 & 1.795 & 0.040 & 1 & 28 &    & < 3.9 & FR0g \\
10C J104954+570456 & 10:49:54.214 & +57:04:56.070 & 7.511 & 0.026 & 7.558 & 0.052 & 1 & 28 &    & < 3.9 & FR0u \\
10C J105000+585227 & 10:50:00.840 & +58:52:28.320 & 0.453 & 0.016 & 0.464 & 0.032 & 1 & 20 &    & < 3.7 & FR0g \\
10C J105007+572020 & 10:50:08.016 & +57:20:18.180 & 0.989 & 0.020 & 1.051 & 0.040 & 1 & 25 &    & < 4.0 & FR0u \\
10C J105007+574251 & 10:50:07.665 & +57:42:49.770 & 0.740 & 0.042 & 0.591 & 0.074 & 1 & 23 &    & < 3.6 & FR0u \\
10C J105009+570724A & 10:50:10.441 & +57:07:25.260 & 0.410 & 0.021 & 0.560 & 0.049 & 1 & 26 &    & < 4.5 & FR0u \\
10C J105009+570724B & 10:50:09.280 & +57:07:16.860 & 0.718 & 0.039 & 0.732 & 0.020 & 1 & 26 &    & < 2.3 & FR0u \\
10C J105020+574048 & 10:50:20.526 & +57:40:46.010 & 0.176 & 0.025 & 0.821 & 0.224 & 3 & 24 & 0.147 & 8.0 & FRII \\
10C J105028+574522 & 10:50:30.016 & +57:45:23.950 & 0.181 & 0.009 & 0.164 & 0.016 & 1 & 13 &    & < 3.6 & FR0u \\
10C J105034+572922 & 10:50:34.180 & +57:29:22.210 & 1.120 & 0.019 & 1.062 & 0.036 & 1 & 29 &    & < 3.5 & FR0u \\
10C J105039+572339 & 10:50:39.560 & +57:23:36.620 & 3.303 & 0.022 & 3.397 & 0.043 & 1 & 26 &    & < 3.6 & FR0c \\
10C J105039+574200 & 10:50:39.116 & +57:41:57.990 & 0.730 & 0.020 & 0.669 & 0.037 & 1 & 28 &    & < 3.5 & FR0u \\
10C J105039+585118 & 10:50:38.693 & +58:51:21.730 & 0.415 & 0.026 & 0.596 & 0.065 & 1 & 27 &    & < 4.0 & FRII \\
10C J105040+573308 & 10:50:40.693 & +57:33:08.330 & 0.931 & 0.020 & 0.913 & 0.039 & 1 & 27 &    & < 3.5 & FR0u \\
10C J105042+575233 & 10:50:42.952 & +57:52:30.010 & 1.433 & 0.044 & 4.210 & 0.187 & 2 & 26 &    & 10.5 & FRI \\
10C J105050+580200 & 10:50:50.636 & +58:01:59.080 & 3.347 & 0.022 & 3.408 & 0.043 & 1 & 25 &    & < 3.5 & FR0u \\
10C J105053+583233 & 10:50:53.980 & +58:32:33.170 & 20.955 & 0.139 & 21.214 & 0.266 & 1 & 60 &    & < 3.4 & FR0u \\
10C J105054+580943 & 10:50:54.103 & +58:09:41.810 & 1.526 & 0.021 & 1.535 & 0.040 & 1 & 27 &    & < 3.4 & FR0u \\
10C J105058+573356 & 10:50:58.379 & +57:33:55.730 & 0.164 & 0.019 & 0.174 & 0.037 & 1 & 24 &    & < 3.1 & FR0g \\
10C J105104+574456 & 10:51:04.674 & +57:44:54.050 & 1.665 & 0.022 & 1.666 & 0.039 & 1 & 29 &    & < 2.6 & FR0g \\
10C J105104+575415 & 10:51:04.570 & +57:54:12.760 & 6.937 & 0.045 & 7.061 & 0.081 & 1 & 28 &    & < 2.4 & FR0u \\
10C J105107+575752 & 10:51:06.606 & +57:57:48.010 & 1.104 & 0.046 & 3.790 & 0.115 & 2 & 28 &    & 8.0 & FRI \\
10C J105115+573552 & 10:51:14.974 & +57:35:51.970 & 0.221 & 0.017 & 0.314 & 0.040 & 1 & 20 &    & < 2.4 & FR0c \\
10C J105121+582648 & 10:51:20.959 & +58:26:51.810 & 1.038 & 0.048 & 3.642 & 0.213 & 3 & 29 & 0.698 & 13.0 & FRII \\
10C J105122+570854 & 10:51:22.051 & +57:08:54.410 & 1.564 & 0.024 & 1.551 & 0.042 & 1 & 28 &    & < 2.5 & FR0c \\
10C J105122+584136 & 10:51:22.056 & +58:41:36.090 & 1.122 & 0.025 & 1.485 & 0.054 & 1 & 30 &    & < 2.4 & FR0u \\
10C J105122+584409 & 10:51:21.471 & +58:44:06.240 & 1.700 & 0.277 & 2.890 & 0.770 & 1 & 268 &    & < 3.6 & FR0u \\
\hline\end{tabular}
\end{table*}

\begin{table*}
\contcaption{}
\small\rm
\medskip
\centering
\renewcommand{\tabcolsep}{1.3mm}
\begin{tabular}{cccddddcdddc}\hline
10C ID & RA & Dec & \dhead{$S_\textrm{peak}$} & \dhead{$\Deltaup S_\textrm{peak}$} & \dhead{$S_\textrm{int}$} & \dhead{$\Deltaup S_\textrm{int}$} & \dhead{$N_\textrm{comp}$} & \dhead{rms / $\muup$Jy} & \dhead{$S_\textrm{core}$}  & \dhead{$\theta$} & FR\\
       &    &     & \dhead{/ mJy}             & \dhead{/ mJy} 					  & \dhead{/ mJy} & \dhead{/ mJy} & & \dhead{beam$^{-1}$} & \dhead{/ mJy} & \dhead{/ arcsec} & class \\
(1) & (2) & (3) & \dhead{(4)} & \dhead{(5)} & \dhead{(6)} & \dhead{(7)} & (8) & (9) & \dhead{(10)} & \dhead{(11)} & (12)\\\hline
10C J105128+570901 & 10:51:28.014 & +57:09:00.630 & 1.261 & 0.029 & 2.307 & 0.077 & 1 & 26 &    & 2.4 & FRII \\
10C J105132+571114 & 10:51:32.329 & +57:11:14.940 & 2.219 & 0.030 & 2.584 & 0.060 & 1 & 28 &    & < 2.4 & FRII \\
10C J105136+572944 & 10:51:36.947 & +57:29:39.850 & 0.643 & 0.026 & 0.813 & 0.057 & 1 & 30 &    & < 2.7 & FR0u \\
10C J105138+574957 & 10:51:38.106 & +57:49:56.610 & 0.182 & 0.035 & 0.181 & 0.061 & 1 & 25 &    & < 2.1 & FR0u \\
10C J105139+580757 & 10:51:38.934 & +58:08:03.120 & 0.495 & 0.041 & 3.640 & 0.346 & 3 & 24 & 0.278 & 18.5 & FRI \\
10C J105142+573447 & 10:51:42.064 & +57:34:47.810 & 0.826 & 0.027 & 0.825 & 0.051 & 1 & 29 &    & < 2.8 & FR0u \\
10C J105142+573557 & 10:51:42.114 & +57:35:54.470 & 2.205 & 0.026 & 2.462 & 0.046 & 2 & 27 & 2.113 & 12.0 & FRII \\
10C J105148+573245 & 10:51:44.150 & +57:33:14.320 & 0.650 & 0.025 & 1.632 & 0.094 & 2 & 18 & 0.483 & 45.0 & FRII \\
10C J105206+574111 & 10:52:06.400 & +57:41:11.780 & 3.670 & 0.027 & 3.905 & 0.049 & 1 & 29 &    & < 2.5 & FR0c \\
10C J105215+581627 & 10:52:16.875 & +58:16:26.960 & 0.979 & 0.024 & 1.051 & 0.046 & 1 & 30 &    & < 2.4 & FR0u \\
10C J105220+585051 & 10:52:21.823 & +58:50:51.670 & 1.529 & 0.027 & 1.974 & 0.059 & 1 & 32 &    & < 2.4 & FR0u \\
10C J105225+573323 & 10:52:25.666 & +57:33:22.610 & 0.265 & 0.019 & 0.630 & 0.068 & 1 & 21 &    & 4.0 & FR0/I \\
10C J105225+575507 & 10:52:25.430 & +57:55:07.620 & 23.689 & 0.083 & 25.599 & 0.154 & 1 & 66 &    & < 2.4 & FR0u \\
10C J105237+573058 & 10:52:36.402 & +57:30:46.600 & 1.126 & 0.049 & 3.642 & 0.251 & 3 & 34 & 0.611 & 39.0 & FRII \\
10C J105240+572322 & 10:52:41.476 & +57:23:20.510 & 0.675 & 0.017 & 0.796 & 0.035 & 1 & 18 &    & < 2.4 & FR0u \\
10C J105243+574817 & 10:52:43.340 & +57:48:13.960 & 1.304 & 0.023 & 1.305 & 0.040 & 1 & 25 &    & < 2.3 & FR0u \\
10C J105327+574546 & 10:53:26.310 & +57:45:43.640 & 0.400 & 0.035 & 1.922 & 0.389 & 2 & 29 & 0.362 & 11.5 & FRII \\
10C J105341+571951 & 10:53:40.880 & +57:19:53.040 & 0.555 & 0.014 & 0.556 & 0.026 & 1 & 18 &    & < 2.3 & FR0u \\
10C J105342+574438 & 10:53:42.190 & +57:44:38.110 & 2.079 & 0.024 & 2.155 & 0.044 & 1 & 27 &    & < 2.3 & FR0/I \\
10C J105400+573324 & 10:54:00.448 & +57:33:21.480 & 0.595 & 0.019 & 0.603 & 0.034 & 1 & 25 &    & < 2.3 & FR0c \\
10C J105425+573700 & 10:54:26.300 & +57:36:48.750 & 1.231 & 0.051 & 34.321 & 2.054 & 5 & 81 & 1.372 & 63.0 & FRII \\
10C J105437+565922 & 10:54:37.875 & +56:59:20.660 & 0.177 & 0.006 & 4.145 & 0.729 & 1 & 28 &    & 31.2 & SFG \\
10C J105441+571640 & 10:54:42.141 & +57:16:39.440 & 0.559 & 0.041 & 0.836 & 0.114 & 2 & 28 &    & 16.0 & FRII \\
10C J105510+574503 & 10:55:10.088 & +57:45:04.540 & 0.564 & 0.023 & 0.592 & 0.043 & 1 & 24 &    & < 2.3 & FR0u \\
10C J105515+573256 & 10:55:16.152 & +57:32:58.620 & 0.188 & 0.033 & 0.611 & 0.176 & 2 & 19 &    & 10.5 & FRI \\
10C J105520+572237 & 10:55:20.970 & +57:22:39.770 & 0.420 & 0.017 & 0.505 & 0.035 & 1 & 20 &    & < 2.4 & FR0u \\
10C J105527+571607 & 10:55:27.593 & +57:16:08.280 & 0.429 & 0.025 & 0.922 & 0.087 & 1 & 28 &    & 3.0 & FR0/I \\
10C J105535+574636 & 10:55:35.291 & +57:46:35.340 & 3.014 & 0.023 & 3.069 & 0.041 & 1 & 27 &    & < 2.3 & FR0u \\
10C J105550+570407 & 10:55:51.150 & +57:04:02.660 & 0.928 & 0.022 & 0.968 & 0.040 & 1 & 26 &    & < 2.4 & FR0u \\
10C J105604+570934 & 10:56:04.753 & +57:09:32.940 & 5.809 & 0.040 & 6.145 & 0.073 & 1 & 28 &    & < 2.4 & FR0c \\
10C J105627+574221 & 10:56:27.725 & +57:42:22.130 & 1.052 & 0.023 & 0.996 & 0.038 & 1 & 26 &    & < 2.2 & FR0u \\
10C J105653+580342 & 10:56:53.117 & +58:03:42.430 & 2.756 & 0.024 & 2.767 & 0.042 & 1 & 26 &    & < 2.3 & FR0g \\
10C J105716+572314 & 10:57:16.144 & +57:23:11.370 & 3.954 & 0.041 & 6.810 & 0.094 & 2 & 28 &    & 8.5 & FRI/II \\
\hline
\end{tabular}
\begin{flushleft}
Notes:\\
(1) source name from the 10C catalogue;
(2) and (3) right ascension and declination of peak flux in VLA image (J2000);
(4) peak flux density in VLA image;
(5) uncertainty in (4);
(6) integrated flux density of source in VLA image;
(7) uncertainty in (6);
(8) number of components source is resolved into;
(9) local rms noise in VLA image, measured in a $100 \times 100$ pixel box away from the source;
(10) flux density of core component. Left blank if core is not distinguishable from the extended emission;
(11) angular size of source measured at 15~GHz (see Section~\ref{section:properties}). Upper limits are marked with `$<$'.
(12) Morphological classification.
\end{flushleft}
\end{table*}

\section{Multi-wavelength counterparts}\label{section:opt-matching}

This sample was matched to the SERVS Data Fusion multi-wavelength catalogue in \citetalias{2015MNRAS.453.4244W}, where we found that 80 out of the 95 sources had a multi-wavelength counterpart. Here we review these matches with the additional information provided by the new, higher resolution 15-GHz VLA data. In order to check the original matching process, we overlaid contours of the new VLA images onto optical images of the area around each source, along with the positions of entries in the SERVS Data Fusion catalogue.

Eight sources were classified as `confused' when matching in \citetalias{2015MNRAS.453.4244W}, as there were several possible optical counterparts within the $3\sigma$ (where $\sigma =$ local rms noise) radio contours and we were unable to decide upon the correct match. With the higher resolution VLA data, we are now able to identify the multi-wavelength counterpart for all eight of these sources. Example images of two of these sources are shown in Appendix~\ref{app:matching}. One previously-confused source, 10CJ105009+570724, is resolved into two separate components in the new data, each with its own separate optical counterpart (see Fig.~\ref{fig:10CJ105009570724}). We therefore split this source into two separate entries in the updated catalogue. This source was previously classified as an FRII source, and is now classified as two separate FR0 sources. Both of these sources have integrated flux densities at 15 GHz greater than 0.5~mJy, the limit originally used to select the sample.

For source 10CJ105327+574546, with the additional information provided by the higher resolution radio data, we now believe the original match is incorrect. This is discussed in more detail in Appendix~\ref{app:matching}. 
Additionally, two sources which were originally classified as not having a match now have a counterpart in the multi-wavelength catalogue. 
For the remaining 84 sources the new data confirm the original classification.

In summary, there are now 95 sources in the sample. Of the original 96 sources, one has been removed (as we believe it is a false detection), and one has been split into 2 sources, adding one source. A further two have been combined into one source as we now believe they are components of the same radio galaxy.
89 out of 95 sources now have a multi-wavelength counterpart. There are 11 original sources with a new match: 10 which previously did not have a match (8 `confused' and 2 `no match') and one where the previous match was incorrect. One of these original source is actually 2 separate sources, both of which have a match, meaning there are a total of 12 sources with a new match.

\subsection{Updated redshifts}

Spectroscopic redshifts are available for four of the new optical counterparts. For the remaining eight new matches, photometric redshift values are used. These are assigned in the same way as described in \citetalias{2015MNRAS.453.4244W}. Four sources have photometric redshifts in the \citet{2013MNRAS.428.1958R} (RR13) catalogue so these values are used, and for the remaining sources photometric redshifts are estimated using the publicly available \textsc{Le Phare} code \citep{1999MNRAS.310..540A,2006A&A...457..841I,2011ascl.soft08009A}; see Section 4 of \citetalias{2015MNRAS.453.4244W} for full details of this process.

Additionally, spectroscopic redshifts are now available for 5 more objects which only had photometric redshift estimates when \citetalias{2015MNRAS.453.4244W} was published. We therefore now have spectroscopic redshifts for 33 sources.

\subsection{Updated HERG/LERG classification}

The 12 sources with new multi-wavelength counterparts are classified as high-excitation or low-excitation radio galaxies in the same way as described in \citetalias{2016MNRAS.462.2122W}. Three of these sources have optical spectra available, so information from these spectra is used to classify them. For the remaining nine sources a combination of mid-infrared colour--colour diagrams, optical compactness and X-ray information is used to classify the sources in the same way as outlined in Section 3 of \citetalias{2016MNRAS.462.2122W}. This resulted in one new HERG and nine new LERGs, the remaining two sources with a new match did not have enough information available to be classified. 

Eight sources which were classified as HERGs or LERGs based on their mid-infrared colours/optical compactness/X-ray information in \citetalias{2016MNRAS.462.2122W} now have optical spectra available, providing an additional check on these classifications. For seven of these eight sources the classifications from the optical spectra confirm the classifications given in \citetalias{2016MNRAS.462.2122W} (3 HERGs and 4 LERGs). One source (10CJ105142+573447), however, was classified as a HERG based on its X-ray emission in \citetalias{2016MNRAS.462.2122W} but its optical spectrum clearly resembles that of a low-excitation galaxy. \citet{2014MNRAS.440..269M} found that 4 of the 46 sources in their sample are classified as LERGs based on their optical emission lines but display HERG-like behaviour in the X-ray.
As classifications based on optical spectra take precedence in our classification scheme, we have changed the classification of this source from HERG to LERG in the updated catalogue.

The number of sources with the different classifications are therefore as follows: 32 HERGs, 43 LERGs and 20 unclassified sources (95 total). A table summarising the properties of the multi-wavelength counterparts and redshift values of all the sources in this sample is available online as supplementary material. A description of the table is given in Appendix~\ref{app:hosts}.

\section{Comparison with the 10C survey}\label{section:10C}

\begin{figure}
\centerline{\includegraphics[width=\columnwidth]{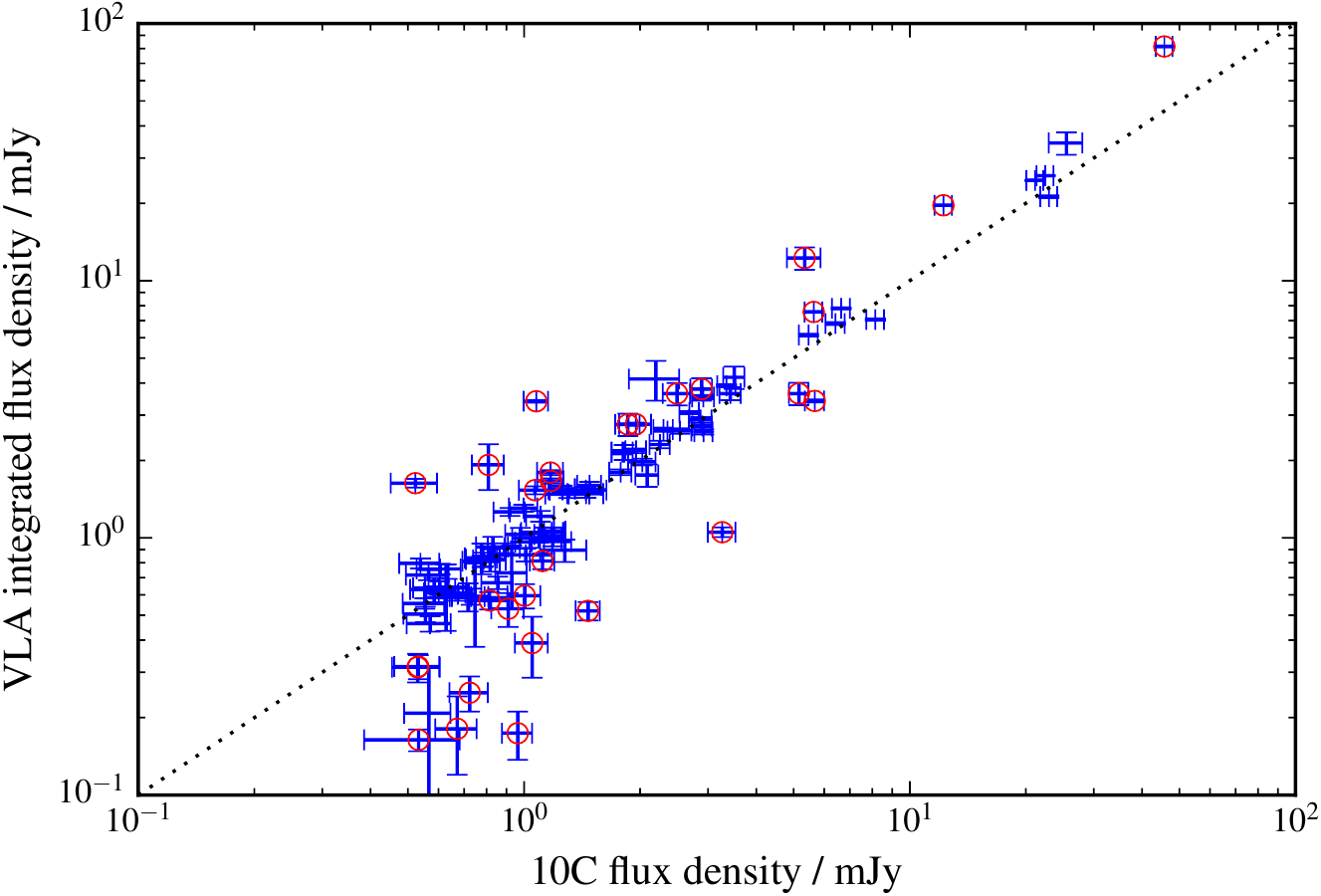}}
\caption{Comparison of flux densities in the original 10C catalogue and the new VLA observations. Integrated flux densities are used for the VLA observations, and in the 10C observations peak flux density is used for unresolved sources and integrated flux density for resolved sources, in the same way as in the 10C survey (see \citealt{2011MNRAS.415.2708D}). Sources where the two values differ by more than $3\sigma$ are marked by red circles.}\label{fig:flux_comparison}
\end{figure}

The observations contributing to the 10C survey were taken between August 2008 and June 2010 and have a synthesised beam size of 30~arcsec, while the VLA observations were taken in May and June 2017 and have a synthesised beam size of $\sim$2~arcsec. Both sets of observations are at 15~GHz, so any differences in the flux densities between the two sets of observations could be due to: 1) uncertainties in the measurements, 2) the different resolutions of the observations, or 3) intrinsic source variability. 

The flux densities of the sources in the 10C survey and the new VLA observations are compared in Fig.~\ref{fig:flux_comparison}. Note that for the purposes of comparing the two surveys sources 10CJ105009+570724A and 10CJ105009+570724B are combined as they cannot be separated in the 10C data. Although the values from the two sets of observations are broadly consistent for the majority of sources, for nearly one third of the sources the values differ by more than $3\sigma$, suggesting that another factor besides the uncertainties in the measurements is contributing towards these differences.

Differences in the measured flux densities between the two sets of observations could be due to the different beam sizes and array configurations of the AMI telescope and the VLA in C configuration. It is possible that diffuse emission from the sources may have a surface brightness too low to be detected in the VLA observations, or be resolved out, resulting in a measured VLA flux density which is lower than the measured 10C flux density. Low-surface brightness emission is clearly more likely to be missed at low signal-to-noise, so this may well explain the large positive flux ratios seen for several sources at low signal-to-noise. This suggests that although these sources appear unresolved in both the 10C and VLA observations, they may have some extended emission on scales between 2 and 30 arcsec. 14 unresolved sources have a 10C flux density which is larger than the VLA flux density by more than $3\sigma$, so may have some emission on these scales.
There is evidence for this from lower-frequency observations of the 10C sample; two of the 14 unresolved sources with 10C flux densities which are larger by more than $3\sigma$ are extended in either the GMRT (610~MHz, resolution 6~arcsec, \citealt{2008MNRAS.387.1037G,2010BASI...38..103G}) or WSRT (1.4~GHz, resolution 11~arcsec, \citealt{2018MNRAS.481.4548P}) observations (see \citetalias{2013MNRAS.429.2080W} for further details of these observations). A further four sources which are compact in both the 10C and VLA observations have extended emission detected at lower frequencies, the flux densities of these sources from the two sets of observations agree within $3\sigma$.

Alternatively, other nearby sources in the AMI beam may be contributing to the measured flux, causing it to be larger in the 10C observations than in the new observations. However, if a source were bright enough to make a significant contribution to the measured flux of a nearby source we would expect it to be detected in the new VLA observations. In one case, 10CJ105009+570724, the 10C detection clearly has contributions from two separate sources as already discussed, both of which are detected in the VLA observation. None of the other sources with higher flux densities in the 10C observations than the VLA observations show evidence of nearby sources in the VLA image.

\begin{figure}
\centerline{\includegraphics[width=\columnwidth]{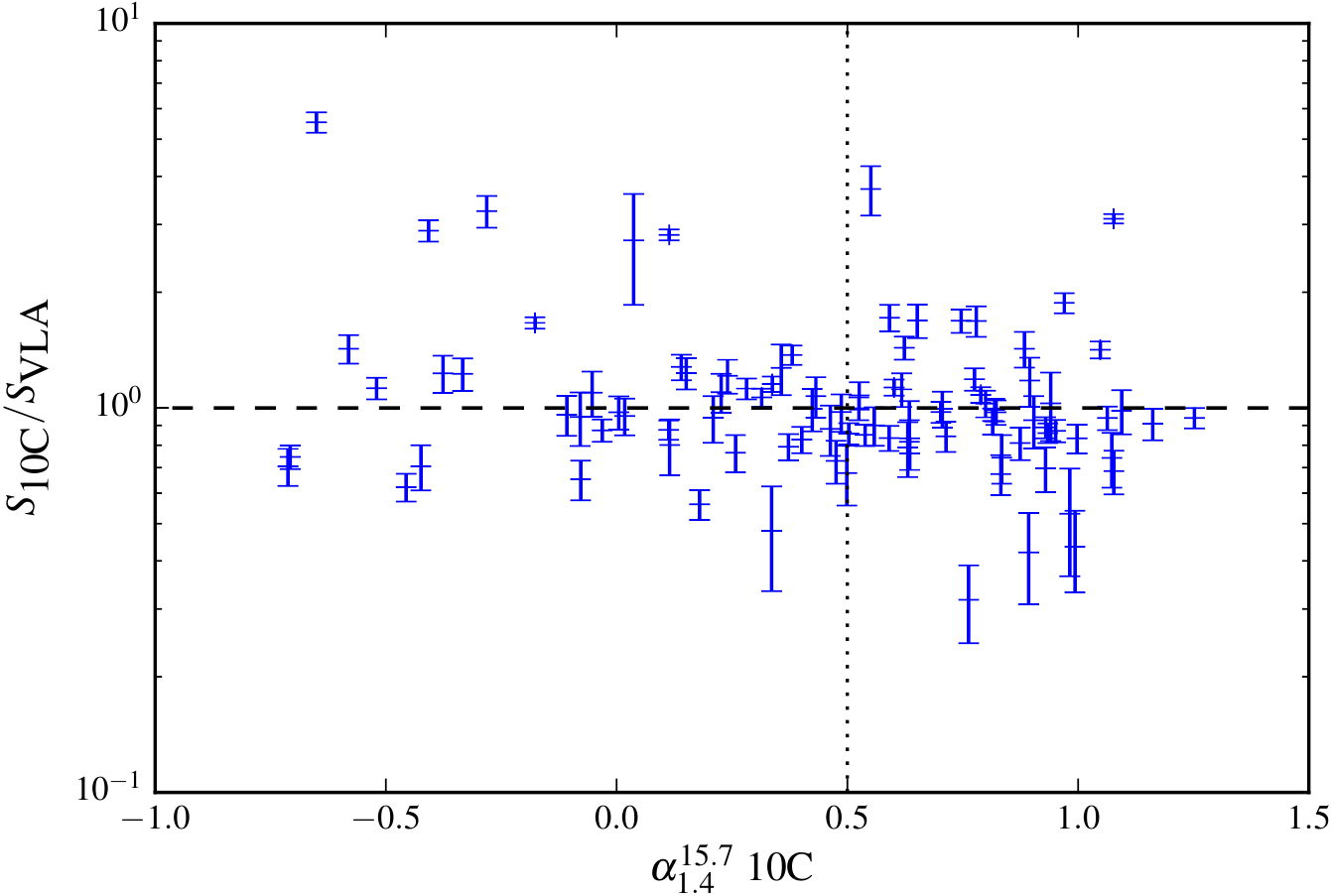}}
\caption{Ratio of flux densities in the 10C and VLA images as a function of spectral index. For the 10C flux densities, peak flux densities are used for sources which appear point-like in the 10C observations, and integrated flux densities are used otherwise, in the same way as in \citet{2011MNRAS.415.2699F}. For the VLA data, integrated flux densities are used for all sources. The dashed line indicates where the ratio of the flux densities from the two surveys is equal to one, and the dotted line is at $\alpha = 0.5$, the standard value used to classify sources as have either steep ($\alpha > 0.5$) or flat ($\alpha < 0.5$) spectra. Source 10CJ105009+570724, which has been split into two, has been recombined for the purpose of this comparison as the two sources are not resolved separately in the 10C observations.}\label{fig:10Cratio_alpha}
\end{figure}

There was an interval of $\sim9$ years between the two sets of observations, so the flux density differences seen could be due to variability. While there have been no studies of the variability of 15-GHz sources in our flux density range, \citet{2006MNRAS.370.1556B} studied the variability of 51 9C sources with $S_{15~\rm GHz} > 25$~mJy over a three year period and found that 29 per cent of sources showed some variability. \citet{2014MNRAS.438..796S} found that one third of the FR0 sources in their sample of nearby 20-GHz selected sources varied by more than 10 per cent over a time period of between 10 and 15 years. We therefore expect some variability for the sources in our sample, so this could be the cause of some of the flux density differences seen. 
Flat-spectrum sources are expected to be more variable than those with steep spectra, so if variability was the main reason for the difference in the flux densities between the two sets of observations, we would expect these differences to be larger for sources with flat spectra. Fig.~\ref{fig:10Cratio_alpha} shows the ratio of flux densities as a function of spectral index; the number of steep and flat spectrum sources have flux densities which differ by more than 3$\sigma$ is very similar (15 flat-spectrum sources, defined as having $\alpha < 0.5$, and 16 steep-spectrum sources, with $\alpha > 0.5$). However, four out of the five sources with flux density values which vary by more than $10\sigma$ have flat spectra, suggesting that intrinsic source variability could be the cause of these large differences.

In summary, the differences in the measured flux densities between the two sets of observations displayed by some sources could be due to intrinsic source variability or extended emission on scales between 2 and 30 arcsec not detected in the VLA observations (or a combination of these two factors). Further observations sensitive to emission on scales between 2 and 30~arcsec are required to distinguish between these two scenarios. 

\section{The dominance of core emission in faint radio galaxies}\label{section:S3}

\begin{figure}
\centerline{\includegraphics[width=\columnwidth]{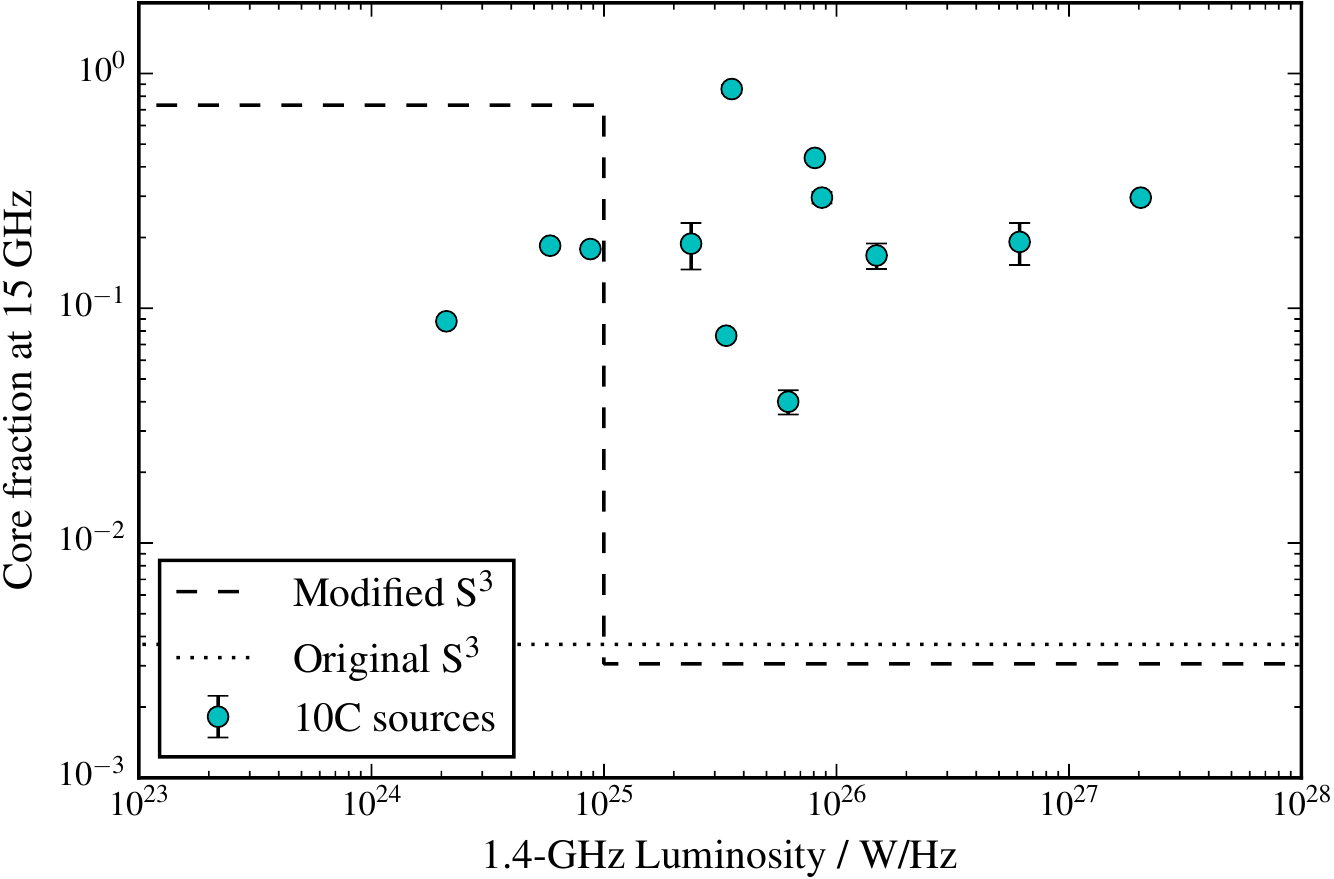}}
\caption{Core fraction at 15 GHz as a function of 1.4-GHz luminosity for extended sources with an identifiable core. The dashed line is the core fraction at 15~GHz used in our proposed modification to the SKA Simulated Skies \citep{2017MNRAS.471..908W} and the dotted line is the mean core fraction in the original simulation. These values are extrapolated to 15~GHz by assuming the spectral index of the core emission and extended emission are 0.0 and 0.75 respectively.}\label{fig:core_frac_L}
\end{figure}

\begin{figure}
\centerline{\includegraphics[width=\columnwidth]{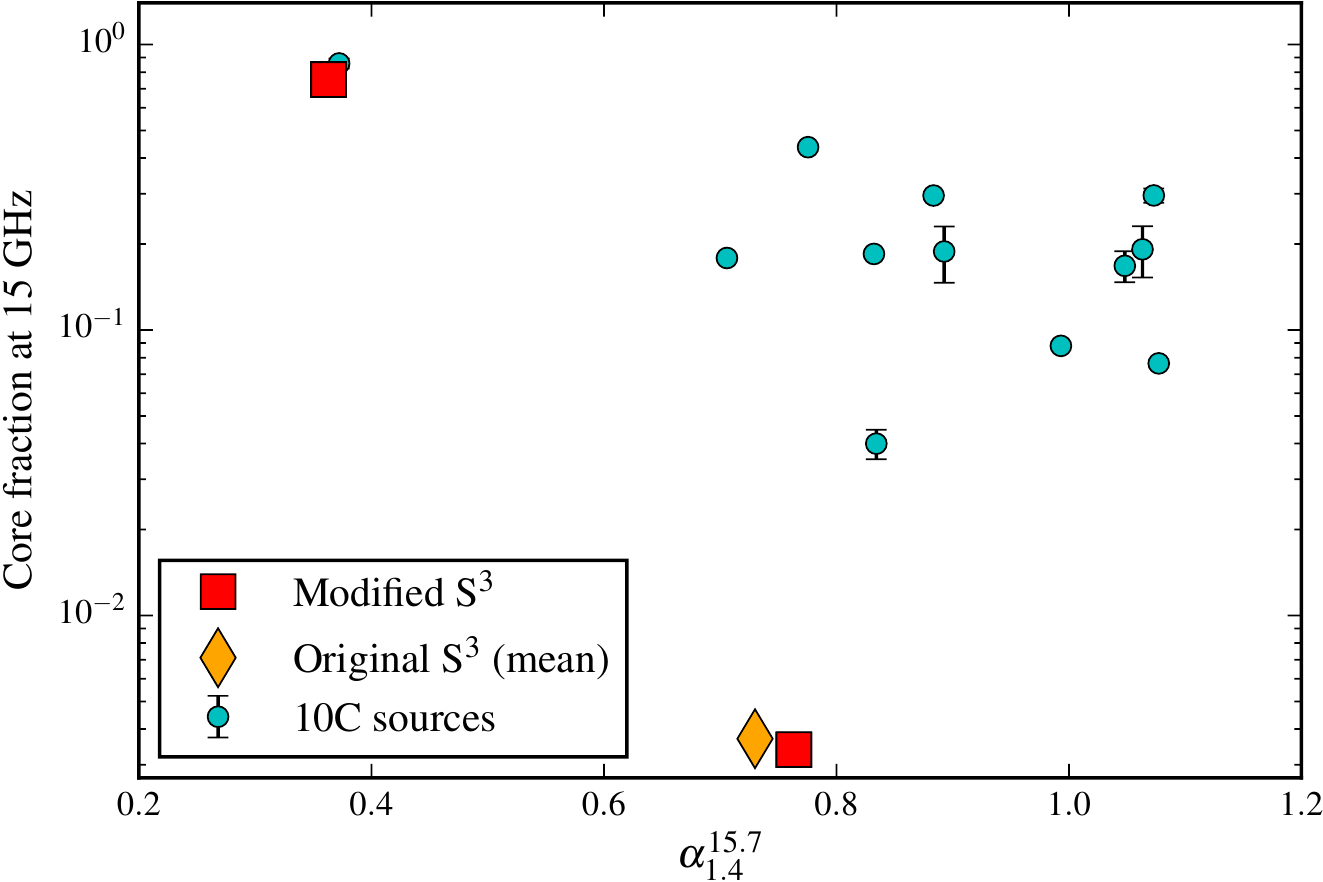}}
\caption{Core fraction at 15~GHz as a function of spectral index between 1.4 and 15 GHz for sources with an identifiable core. The core fractions at 15~GHz used in our proposed modification to the SKA Simulated Skies \citep{2017MNRAS.471..908W} are plotted as squares (red in the online version), and the mean value in the original simulation is shown as a diamond (orange in the online version. These values are extrapolated to 15~GHz by assuming the spectral indices of the core emission and extended emission are 0.0 and 0.75 respectively. }\label{fig:core_frac_alpha}
\end{figure}

While there have been several studies of the core emission in powerful radio galaxies (e.g.\ \citealt{2011MNRAS.417..184B}), the cores of fainter radio galaxies ($S \lesssim 1$~mJy) have not been well studied. In \citet{2017MNRAS.471..908W}, we showed that if the cores of less-powerful ($L_{1.4~\rm GHz} <10^{25} \, \textrm{W Hz}^{-1}$) FRI galaxies are more dominant than assumed in the SKA Simulated Skies (S$^3$; \citealt{2008MNRAS.388.1335W,2010MNRAS.405..447W}), the simulated source counts are a much better fit to observations at higher radio frequencies ($\gtrsim 15$~GHz). 

As described fully in \citet{2017MNRAS.471..908W} and summarised here, we found that the observed source counts at $>15$~GHz can be better reproduced by making the following simple modification to the simulated catalogue. A fraction $x$ of the total 1.4-GHz flux density of each FRI source in the simulation is assumed to be in the core, and assigned a spectral index between 1.4 and 18~GHz of $\alpha = 0$. The remaining flux density is assumed to be extended emission and therefore given a spectral index of 0.75. This is then used to produce a revised flux density for each FRI source at 4.8 and 18~GHz. We found that the best-fit values for $x$ were 0.31 for FRI sources with $L_{1.4~\rm GHz} <10^{25} \, \textrm{W Hz}^{-1}$ and $2.5 \times 10^{-3}$ for sources with $L_{1.4~\rm GHz} <10^{25} \, \textrm{W Hz}^{-1}$. This proposed modification is clearly a simplification, and for example does not include effects such as spectral ageing; however it demonstrates that the high-frequency source counts can be better reproduced by assuming that the FRI sources in the simulation with $L_{1.4~\rm GHz} <10^{25} \, \textrm{W Hz}^{-1}$ are more core dominated than previously assumed in the model used to generate the simulation. The observations described in this paper allow us to measure the core dominance in this sample of faint, high-frequency selected radio galaxies and therefore test this proposed modification.

Out of 95 sources in the sample, 17 sources are resolved into more than one component and 12 of the 17 have an identifiable core. For these sources we can directly measure the core fraction, $x$, defined as $x = S_{\rm core} / S_{\rm tot}$. 
These core fractions are compared to those from the modified SKA Simulated Skies presented in \citet{2017MNRAS.471..908W} in Figs. \ref{fig:core_frac_L} and \ref{fig:core_frac_alpha}. 
The modified S$^3$ core fractions are average values for the whole population; clearly we expect the values for individual sources to display considerable scatter. Most of the lower-luminosity 10C sources (with $L < 10^{25} \textrm{W Hz}^{-1}$) are not resolved in the VLA observations so cannot be included in this plot. These low-luminosity sources generally have flatter spectra (as demonstrated in \citetalias{2013MNRAS.429.2080W}) and therefore are likely to have higher core fractions.
The low luminosity sources which can be plotted clearly have larger core fractions than assumed in the original simulation, providing support for our proposed modification to the simulation. Note however that the more luminous sources also seem to have larger core factions than assumed in either the original simulation or the modified version.

\begin{figure}
\centerline{\includegraphics[width=\columnwidth]{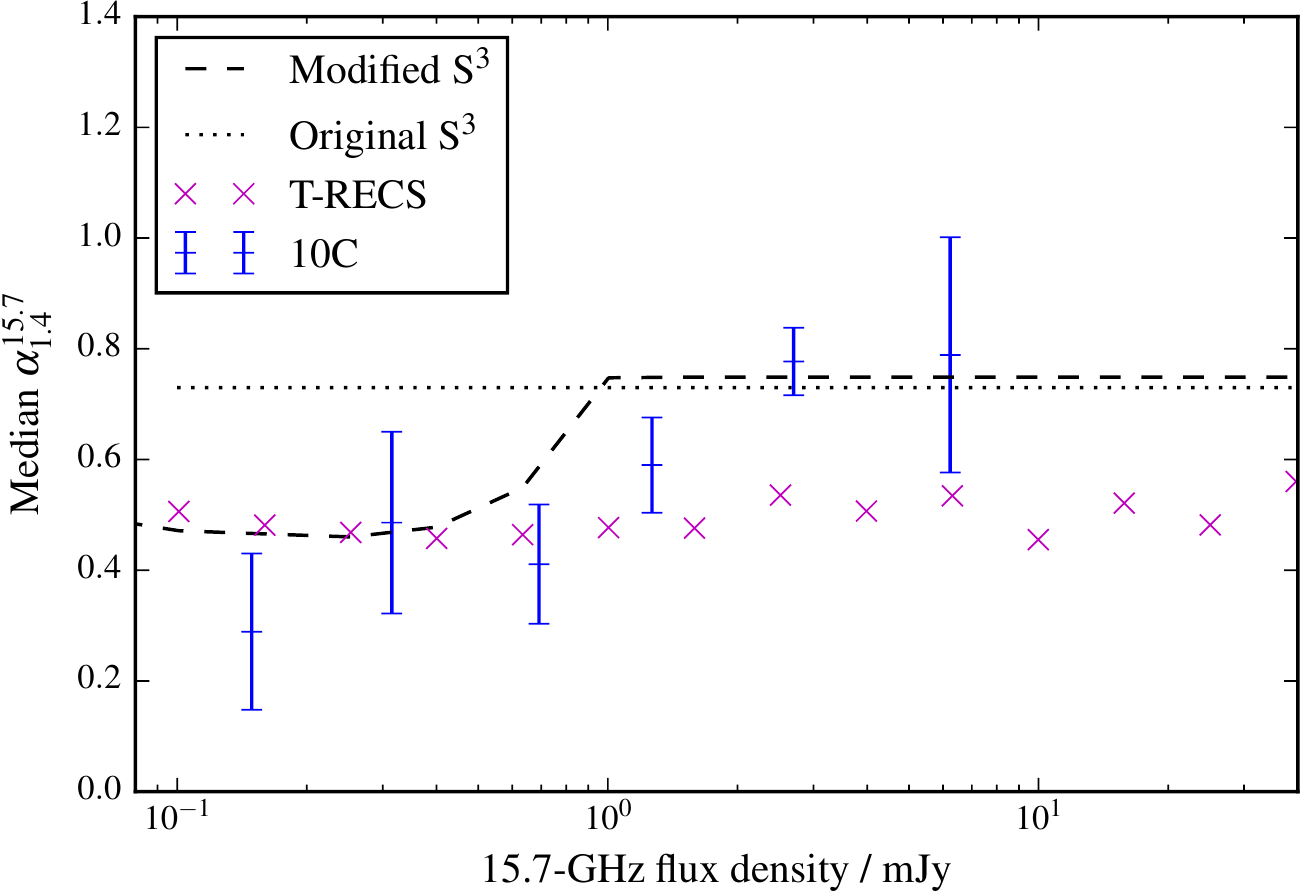}}
\caption{Median spectral index in a series of 15-GHz flux density bins. The median values from the full Lockman Hole 10C sample are shown, along with the original and modified S$^3$ simulation and the T-RECS simulation.}\label{fig:TRECS}
\end{figure}

The newly released T-RECS simulation \citep{2019MNRAS.482....2B} is a much better fit to the observed high-frequency source counts than S$^3$ (see \citeauthor{2019MNRAS.482....2B} Fig. 6). While \citeauthor{2019MNRAS.482....2B} do not provide the information required to calculate core fractions, we are able to compare the spectral indices of the sources in the T-RECS simulated catalogue to those in S$^3$ and in our 10C sample, as shown in Fig.~\ref{fig:TRECS}. Below 1~mJy the T-RECS spectral indices are close to those measured from the 10C sample, which is consistent with the fact that T-RECS is able to reproduce the observed high-frequency source counts, while S$^3$ is not. For the bright sources, however, the T-RECS median spectral indices appear to be lower (i.e.\ the spectra are flatter) than those in the 10C sample.

\begin{figure}
\centerline{\includegraphics[width=\columnwidth]{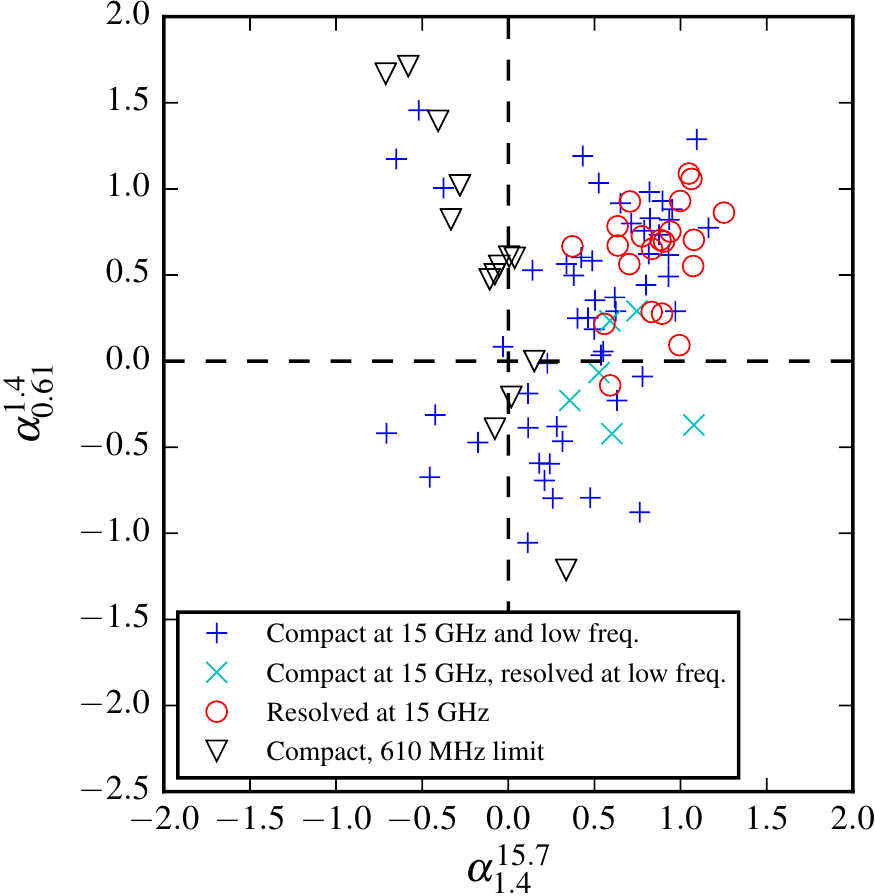}}
\caption{Radio colour--colour diagram. Sources which are resolved and unresolved in the new 15~GHz VLA observations are shown separately. The 6 sources which are unresolved in the new observations are but are extended at 1.4~GHz/610~MHz are shown as cyan crosses. Sources with an upper limit on their 610~MHz flux density are marked with black triangles (these are all unresolved in the 15-GHz and lower-frequency observations). }\label{fig:colour}
\end{figure}

\begin{figure}
\centerline{\includegraphics[width=\columnwidth]{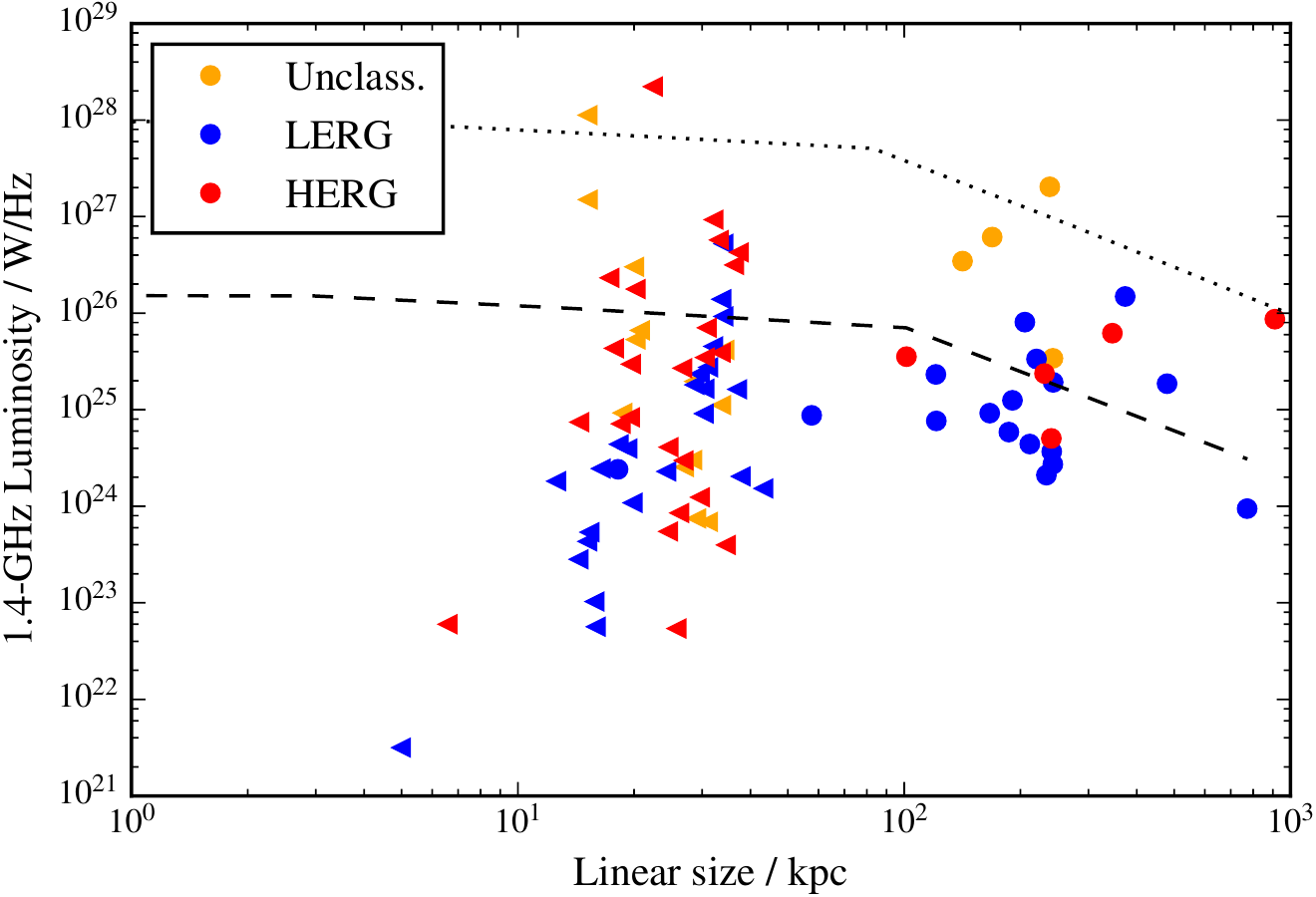}}
\smallskip
\centerline{\includegraphics[width=\columnwidth]{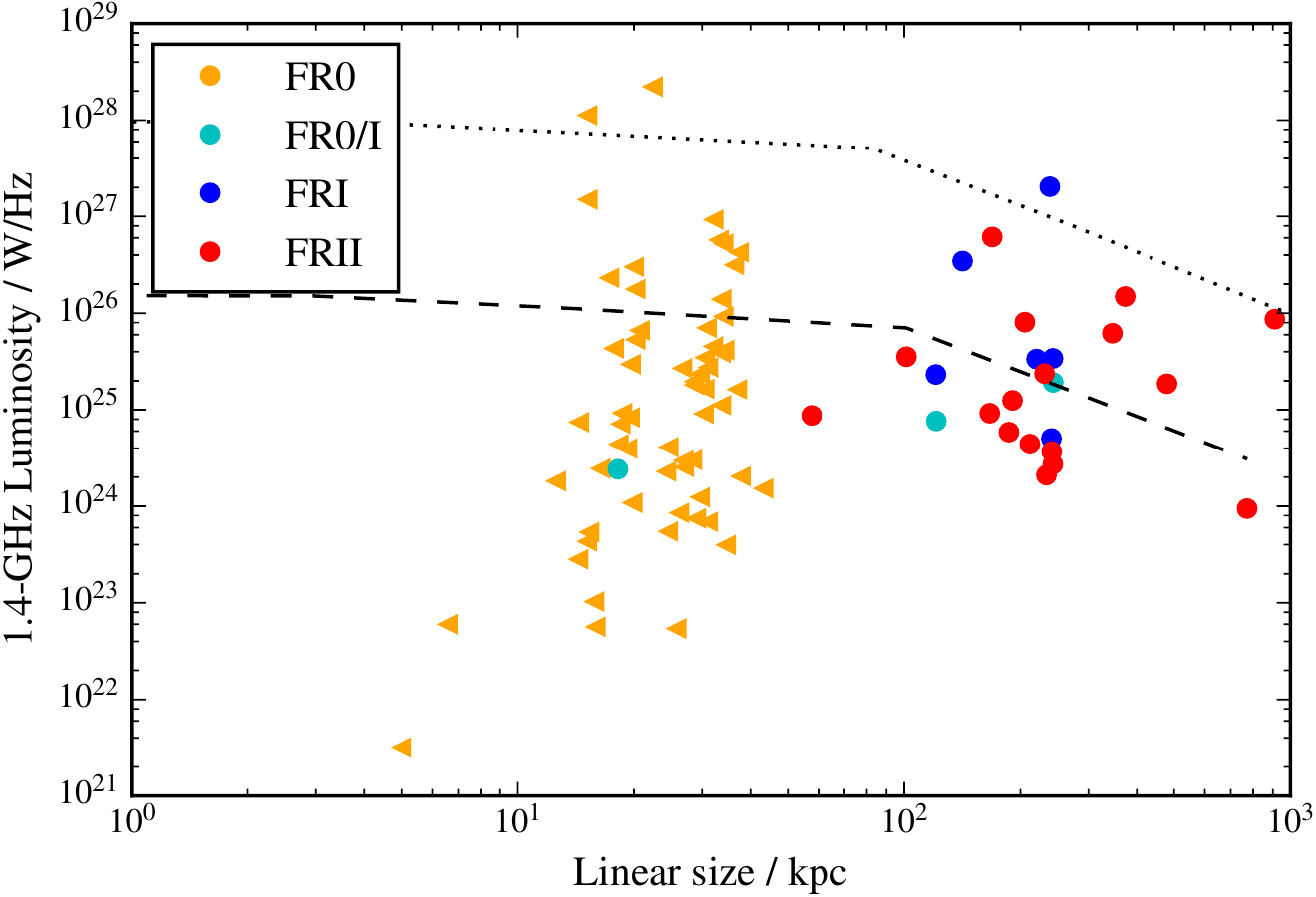}}
\caption{Radio power -- linear size diagram. The lines are example evolutionary tracks for FRI (dashed line) and FRII (dotted line) sources from \citet{2012ApJ...760...77A}. For the compact sources sizes are measured from the VLA observations and for the extended sources they are taken from the lower-frequency observations. Triangles indicate upper limits on linear sizes. In the top panel HERGs, LERGs and unclassified sources are shown separately. In the bottom panel sources are coloured according to their FR morphological classification; these classifications are based on 1.4~GHz/610~MHz observations and are described in \citetalias{2016MNRAS.462.2122W}.}\label{fig:sizelum}
\end{figure}

The spectral shapes of the sources in our sample are illustrated by the radio colour--colour plot shown in Fig.~\ref{fig:colour}, where the sources which are resolved in the new VLA observations are shown separately. The majority of the resolved sources have steep spectra (as expected), while the unresolved sources display a range of different spectral shapes. In particular, a number of compact sources are found in the bottom left and right quadrants, where high-frequency peakers and GPS sources respectively are expected to lie. This suggests that these sources could be young radio galaxies.

The 1.4-GHz luminosities of the radio galaxies in this sample are shown in Fig.~\ref{fig:sizelum} as a function of linear size, in which the unresolved compact sources are shown as triangles. The resolved sources all have 1.4-GHz luminosities greater than $10^{24} \textrm{W Hz}^{-1}$, while the unresolved sources span the full range of luminosities $10^{21} < L_{1.4~\rm GHz} / \textrm{W Hz}^{-1} < 10^{29}$.
Evolutionary tracks from \citet{2012ApJ...760...77A} for FRI and FRII sources are shown in Fig.~\ref{fig:sizelum}; the majority of the sources lie below the FRI track, although since most of these sources have upper limits on their linear sizes they could in fact move to the left and lie in the region of the diagram indicative of young sources.
Note that many of the FRII sources in our sample have relatively low luminosities, lying well below the evolutionary track for FRII sources. 

Similar numbers of the compact sources are classified as LERGs and HERGs (23 and 27 out of 67 sources respectively) while a far greater proportion of the extended sources are classified as LERGs than HERGs (19 are LERGs while only 5 are HERGs).

While the extended sources mostly have $L_{1.4~\rm GHz} > 10^{24} \textrm{W Hz}^{-1}$, steep spectra and are LERGs, the compact 10C sources display a range of spectral shapes and radio powers and are a mixture of HERGs and LERGs. This suggests that they are a composite population containing a mixture of different source types. 
31 per cent of the FR0 sources in this sample are candidate GPS or CSS sources, implying that they may be young radio galaxies which are yet to develop powerful extended emission. Alternatively, the compact sources may be intermittent AGN which are currently in an active phase, but this active phase is not sustained for long enough to develop extended radio jets. Another possibility is that the intrinsic properties of the jets (e.g.\ lower bulk speeds) may mean they are more easily disrupted preventing them from forming extended structures. Further study is required to distinguish between these scenarios.

It has been suggested (e.g.\ \citealt{2016AN....337..105S}) that some FR0 sources do have extended radio jets, but the emission is below the surface-brightness limit of most large-scale radio surveys, meaning that these sources appear compact. \citet{2017MNRAS.464.4706S} modelled the emission from three AGN and showed that the surface brightness of their lobes would be too low to be detected by most surveys with the VLA and LOFAR.
This idea is supported by the fact that, as discussed in Section~\ref{section:10C}, there is evidence that some of the unresolved sources in our sample may have emission on scales between 2 and 30~arcsec, the surface brightness of which is too low to be detected in our VLA observations. The MeerKAT radio telescope in South Africa would be the ideal instrument with which to investigate this as the large number of short baselines in the array provide excellent sensitivity to diffuse, low-surface brightness emission. Future work will involve performing such a study using MeerKAT data in the COSMOS field.

The results in this paper show that compact radio galaxies, or those with only very weak extended emission, are the dominant population in the 10C survey (making up 67/95 sources in this sample), rather than FRI sources dominated by extended emission as predicted by S$^3$. Sources with compact morphologies (FR0s) have been found to be the dominant population in lower-powered ($ L_{1.4~\rm GHz} / \rm W \, Hz^{-1} < 10^{25}$) radio-loud AGN samples in the nearby universe ($z < 0.1$). For example, both `FR0CAT' selected from FIRST and NVSS at $z<0.05$ \citep{2018A&A...609A...1B} and the \citet{2014MNRAS.438..796S} study of AT20G-6dFGS sources selected at 20~GHz with $z<0.1$ are dominated by compact radio galaxies. As the median redshift for our sample is 0.9, our results show that this is also the case at higher redshifts. Recent high-resolution VLA observations of 18 nearby FR0 sources by \citet{2019MNRAS.482.2294B} found that 14 were still unresolved on scales $\sim 0.3$~arcsec, constraining their linear sizes to $<1$~kpc. Higher-resolution observations are required to determine if our higher-redshift sources are compact on these scales. 

The compact radio galaxies in our sample are a mixed population; the \citet{2014MNRAS.438..796S} study produced similar results, finding that the compact radio galaxies in their sample display a range of spectral indices and are a mixture of different types of object. Additionally, recent VLBI observations of 14 nearby FR0 sources by \citet{2018ApJ...863..155C} found that their sample is a heterogeneous population consisting of a mixture of compact steep spectrum and GHz-peaked spectrum sources, displaying a range of spectral indices and other source properties.

We have previously compared the properties of 10C sources to those in the AT20G-6dFGS sample (\citetalias{2016MNRAS.462.2122W}), and suggested that the compact radio galaxies in our sample may be higher-redshift analogues of the AT20G-6dFGS sources. Both our sample and the AT20G-6dFGS sample contain a mixture of HERGs and LERGs, while the sources in the \citet{2018A&A...609A...1B} sample are almost exclusively hosted by LERGs. Our sample therefore allows us to investigate the properties of compact HERGs, which have not been as well-studied as the compact LERGs prevalent at lower redshifts. In the local universe most HERGs are FRII sources, so these compact HERGs at higher redshifts could be young sources which will later develop into FRII sources. Some support for this comes from the fact that in the current sample 34 per cent (11/32) of the HERGs are candidate CSS or GPS sources (FR0c or FR0g), providing evidence that these may be young sources, compared to 14 per cent (6/42) of the LERGs. 

The properties of the sources in our sample therefore suggest that these compact, high-frequency selected radio galaxies are a mixed population containing young radio galaxies, older sources which have failed to produce large jets, and, potentially, sources with extended emission with a surface brightness which is too low to be detected in most observations. 


\section{Conclusions}\label{section:conclusions}

We have presented new 15 GHz VLA observations with arcsec resolution of a complete sample of 95 sources selected from the 10C survey at 15.7~GHz. With these higher-resolution observations we were able to identify multi-wavelength counterparts for 8 additional sources, meaning that we now have multi-wavelength counterparts for 89 out of the 95 sources in the sample. The main results of this work are as follows.

\begin{itemize}
	\item 73 out of 95 (77 per cent) of this complete sample of 10C sources appear unresolved in the VLA observations. However, six of these unresolved sources show some extended emission at lower frequencies and are therefore not genuinely compact, meaning 67 out of the 95 sources (71 per cent) are compact on arcsec scales. 

	\item The 10C survey is therefore dominated by compact radio galaxies, or radio galaxies with only very weak extended emission. \citet{2014MNRAS.438..796S} found that compact radio galaxies are the dominant population in high-frequency selected samples in the local universe, so this result shows that this is also the case at $z \sim 1$.

	\item This provides support for our proposed modification to the SKA Simulated Skies \citep{2017MNRAS.471..908W}, where we showed that if the flat-spectrum cores of faint ($L<10^{25} \, \textrm{W Hz}^{-1}$) radio galaxies are assumed to be more dominant than previously thought, the high frequency (18 GHz) source counts and spectral index distributions can be reproduced. However, higher-resolution imaging is required to separate the core emission from any small-scale jet emission and fully constrain the fraction of the emission originating from the core.

	\item The compact radio galaxies in the 10C survey are a mixture of HERGs (40 per cent of compact sources) and LERGs (34 per cent) and display a wide range of spectral indices ($-0.6 < \alpha^{15.7}_{0.61} < 1.2$) and radio powers ($10^{21} < L_{1.4~\rm GHz} / \textrm{W Hz}^{-1} < 10^{29}$), indicating that they are a mixed population. These sources may be the higher-redshift ($z \sim 1$) analogues of the compact FR0 sources found in the local universe ($z \sim 0.1$) studied by \citet{2014MNRAS.438..796S}, \citet{2018ApJ...863..155C} and \citet{2018A&A...609A...1B}.

   \item In contrast, the extended 10C sources generally have steep spectra, $L_{1.4~\rm GHz} > 10^{24} \, \textrm{W Hz}^{-1}$ and are LERGs.
\end{itemize}

Most studies of compact radio galaxies have been at low redshift and are dominated by LERGs. This sample, on the other hand, contains a significant number of compact HERGs and further studies of these objects has the potential to provide insights into the origin and evolution of HERGs.

\section*{Acknowledgements}

The authors thank the anonymous referee for their helpful comments. IHW acknowledges the financial assistance of the South African Radio Astronomy Observatory (SARAO) towards this research (www.ska.ac.za). IHW thanks the South African Astronomical Observatory, where some of this work was carried out. The National Radio Astronomy Observatory is a facility of the National Science Foundation operated under cooperative agreement by Associated Universities, Inc. This research has made use of \textsc{Astropy}, a community-developed core \textsc{Python} package for Astronomy \citep{2013A&A...558A..33A}; \textsc{APLpy}, an open-source plotting package for \textsc{Python} \citep{2012ascl.soft08017R} and NASA's Astrophysics Data System.

%
%

\setlength{\labelwidth}{0pt}

\bsp

\appendix
\section{Details of updated optical counterparts}\label{app:matching}

\begin{figure}
\centerline{\includegraphics[width=\columnwidth]{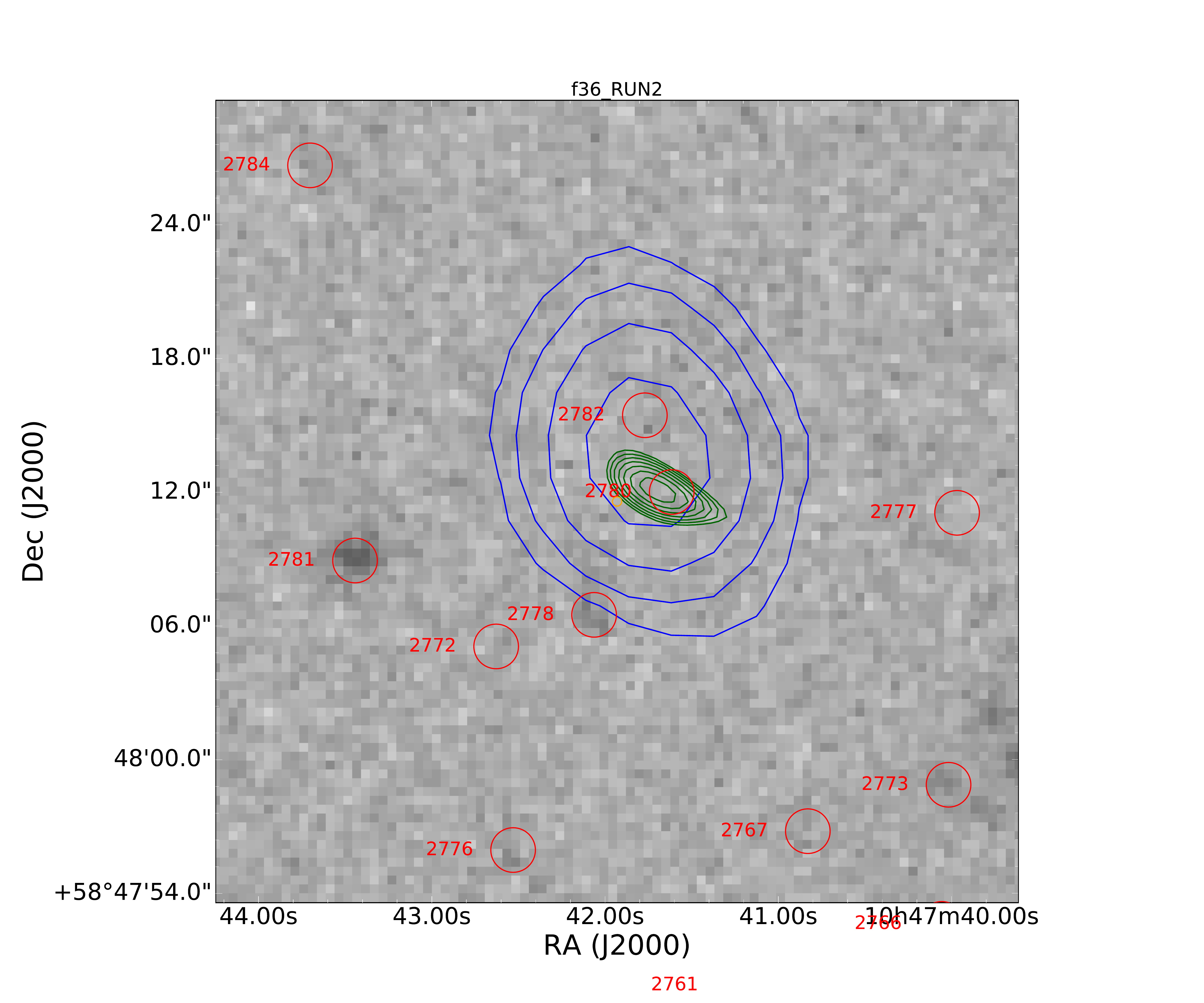}}
\centerline{\includegraphics[width=\columnwidth]{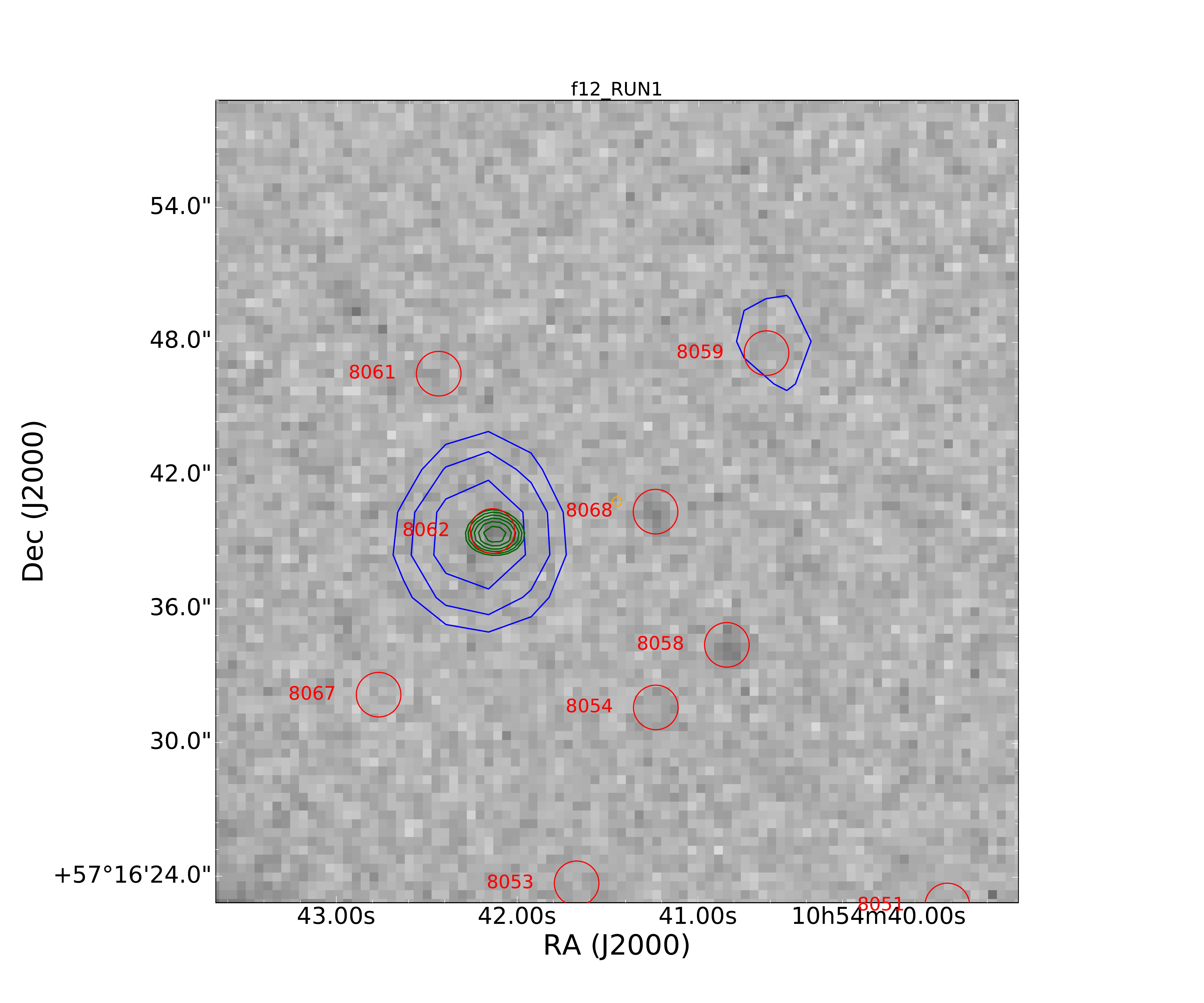}}
\caption{Two sources previously classified as `confused'. The greyscale is the $r$-band optical SDSS image, the blue contours are 1.4 GHz FIRST data and the green contours are the new 15-GHz VLA data. In both cases the contours start at 2$\sigma$, where $\sigma$ is the local rms noise in the relevant radio image. The red circles indicate the positions of sources in the multi-wavelength Data Fusion catalogue. Note that this catalogue is selected in the mid-infrared, which is why objects are not always visible in the optical greyscale image shown here. The images are centred on the position of the source in the original 10C catalogue. The images are $0.3 \times 0.3$ arcmin$^2$.}\label{fig:confused}
\end{figure}

In Section~\ref{section:opt-matching} we reviewed, with the addition of the higher-resolution VLA data, the multi-wavelength counterparts originally assigned to these sources in \citetalias{2015MNRAS.453.4244W}. Some notes on individual sources are presented in this Appendix. 

There are eight sources which were classified as `confused' in \citetalias{2015MNRAS.453.4244W}; we are now able to identify an optical counterpart for all eight sources. Two examples are shown in Fig.~\ref{fig:confused}.

\begin{figure}
\centerline{\includegraphics[width=\columnwidth]{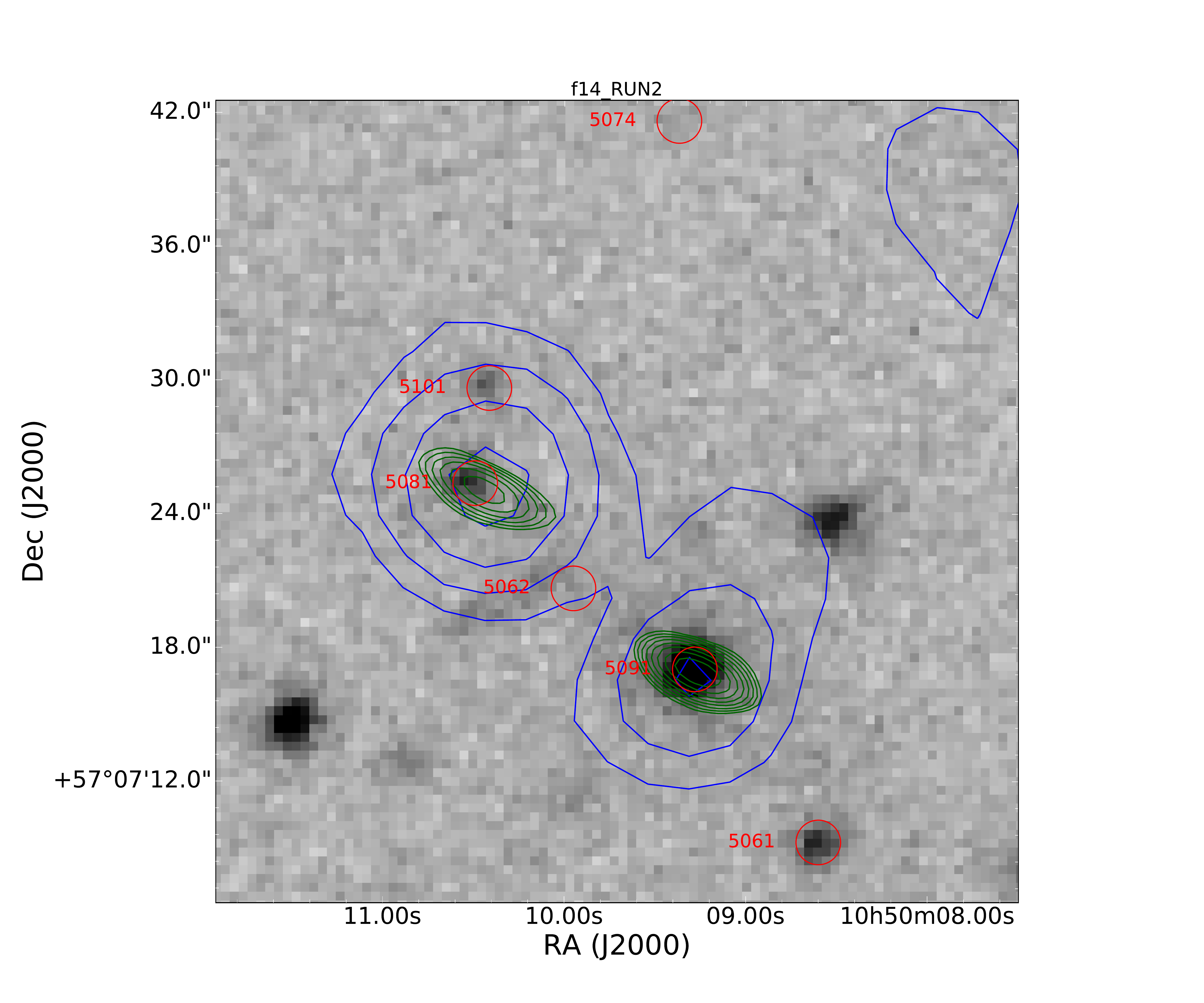}}
\caption{Source 10CJ105009+570724, which was previously classified as confused. We now think this source is 2 separate sources, which match to the objects labelled 5081 and 5091. Contours and labels are the same as in Fig.~\ref{fig:confused}.}\label{fig:10CJ105009570724}
\end{figure}

Source 10CJ105009+570724, which was previously classified as confused, is shown in Figure~\ref{fig:10CJ105009570724}. This source is resolved into two separate components in the VLA image, both of which have a match in the Data Fusion catalogue. We therefore believe that this source is actually two separate objects, and have split it into two entries in the final catalogue (10CJ105009+570724A and 10CJ105009+570724B).

\begin{figure}
\centerline{\includegraphics[width=\columnwidth]{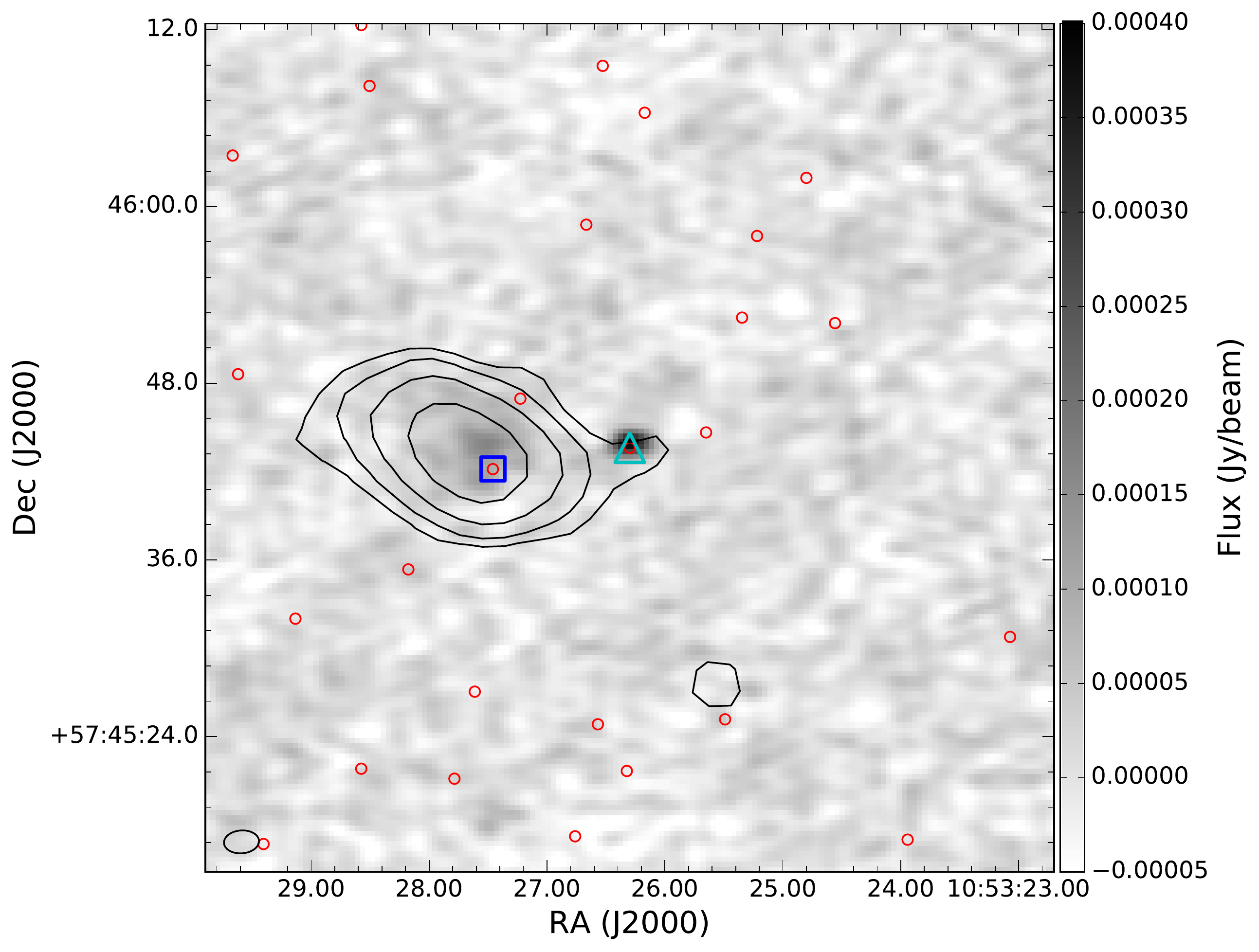}}
\caption{Source 10CJ105327+574546. The background greyscale is the new 15-GHz VLA data and the contours are 610~MHz GMRT data from \citet{2008MNRAS.387.1037G}. The red circles indicate the positions of sources in the multi-wavelength Data Fusion catalogue \citep{2015fers.confE..27V}.  The original match was the object marked with a blue square. With the addition of the new VLA data we now believe that the match is actually the object marked by the cyan triangle, as this is appears to be the core of the source, and the more diffuse emission to the left (which is brighter at 1.4 GHz, indicating that it has a steep spectrum) is a lobe.}\label{fig:wrong}
\end{figure}

Source 10CJ105327+574546 is shown in Fig.~\ref{fig:wrong}. We now believe that the optical counterpart originally assigned to this source is incorrect. Based on the new information from the VLA observations, the correct optical counterpart is actually the object marked with the orange star, as this appears to be coincident with the core of the source. The more diffuse emission to the left is brighter at 1.4 GHz, indicating that it has a steep spectrum, which is consistent with this being one lobe of the radio source.

\section{Host galaxy catalogue}\label{app:hosts}

A table summarising the multi-wavelength data, redshift estimates and HERG/LERG classifications for the sources in the sample discussed in this paper is available online. This is an update of the information provided in Table A1 in \citetalias{2015MNRAS.453.4244W} and Tables 2 and 5 in \citetalias{2016MNRAS.462.2122W} in light of the revised multi-wavelength counterparts discussed in Section~\ref{section:opt-matching} and Appendix~\ref{app:matching}. Table~\ref{tab:columns} provides a summary of the columns in the catalogue, the full catalogue is available online.

\begin{table*}
\caption{The columns of the table summarising the properties of the host galaxies for the sources in this sample, which is available online as supplementary material. This is an update of the information provided in Table A1 in \citetalias{2015MNRAS.453.4244W} and Tables 2 and 5 in \citetalias{2016MNRAS.462.2122W} in light of the revised multi-wavelength counterparts discussed in Section~\ref{section:opt-matching} and Appendix~\ref{app:matching}.}\label{tab:columns}
\bigskip
\begin{tabular}{lp{10cm}}\hline
\vpad
Column heading & Description (if applicable)\\\hline
10C ID & ID from the 10C catalogue\\
$g$ & $g$-band magnitude\\
$i$ & $i$-band magnitude\\
$r$ & $r$-band magnitude\\
$z$ & $z$-band magnitude\\
$J$ & $J$-band magnitude\\
$K$ & $K$-band magnitude\\
SERVS1 / $\muup$Jy & \\
SERVS2 / $\muup$Jy & \\
SWIRE1 / $\muup$Jy & \\
SWIRE2 / $\muup$Jy & \\
SWIRE3 / $\muup$Jy & \\
SWIRE4 / $\muup$Jy & \\
Best $z$ & Final redshift value (see section 4.3 in \citetalias{2015MNRAS.453.4244W}). Null if no redshift value is available.\\
$z$ flag & Origin of final redshift value. 1 = spectroscopic, 2 = \citet{2013MNRAS.428.1958R}, 3 = \citet{2012ApJS..198....1F}, 4 = \textsc{Le Phare}. Null if no redshift value is available.\\
Optical compact & E = source is classified as extended, P = source is classified as point-like (see Section 3.1 of \citetalias{2016MNRAS.462.2122W}). Null if information is not available.\\
Lacy et al. AGN area & Y = source is located inside the \citet{2004ApJS..154..166L} area on the mid-infrared colour--colour diagram, N = source is located outside this region (including sources with limits which must lie outside this region), see Section 3.2 in \citetalias{2016MNRAS.462.2122W}. Null if information is not available.\\
X-ray detection & Y = source is detected by an X-ray survey, N = source lies inside X-ray survey area but is not detected (see Section 3.3 in \citetalias{2016MNRAS.462.2122W}). Null if source is outside the X-ray survey area.\\
Spectral class & H = source is classified as a HERG based on its optical spectrum, L = source is classified as a LERG based on its optical spectrum. (See Section 3.5 in \citetalias{2016MNRAS.462.2122W}). Null if no spectrum is available.\\
HERG class & Overall HERG or LERG classification. Null if there is insufficient information available to classify the source.\\
\hline
\end{tabular}
\end{table*}

\label{lastpage}
\end{document}